\documentstyle[prb,aps]{revtex}
\begin{document}
\def\rhov{{\mbox{\boldmath{$\rho$}}}}
\input{psfig}
\bibliographystyle{unsrt}
%\tightenlines
\draft
\title{QUANTUM THEORY OF CHIRAL INTERACTIONS IN CHOLESTERIC LIQUID CRYSTALS}
\author{S. A. Issaenko, A. B. Harris, and T. C. Lubensky}
\address{Department of Physics,
         University of Pennsylvania,
         Philadelphia, PA 19104}
\maketitle
\begin{abstract}
The effective chiral interaction between molecules arising from
long--range quantum interactions between fluctuating charge moments
is analyzed in terms of a simple model of chiral molecules. This model
is based on the approximations that a) the dominant excited states of a
molecule form a band whose width is small compared to the average
energy of excitation above the ground state and b) biaxial
orientational correlation between adjacent molecules can be neglected.
Previous treatments of quantum chiral interactions have been based on a
multipole expansion of the effective interaction energy within
second--order perturbation theory. We consider a system consisting
of elongated molecules and, although  we invoke the expansion in
terms of coordinates transverse to the long axis of constituent
molecules, we treat the longitudinal coordinate exactly. Such an
approximation is plausible for molecules in real liquid crystals.
The macroscopic cholesteric wave vector ${\bf Q}$ ($Q = 2 \pi /P$,
where $P$ is the pitch) is obtained via $Q=h/K_2$, where $K_2$ is
the Frank elastic constant for twist and $h$ is the torque field
which we calculate from the effective chiral interaction 
$\kappa_{IJ} {\bf a}_I \times {\bf a}_J \cdot {\bf R}_{IJ}$, where
the unit vector ${\bf a}_I$ specifies the orientation of molecule $I$
and ${\bf R}_{IJ}$ is the displacement of molecule $I$ relative to
molecule $J$.  We identify two distinct physical
limits depending on whether one or both of the interacting molecules
are excited in the virtual state.  When both molecules are excited,
we regain the $R_{IJ}^{-8}$ dependence of $\kappa_{IJ}$ on
intermolecular separation
found previously by van der Meer et al. The two--molecule, unlike the
one--molecule term, can be interpreted in terms of a superposition of
pairwise interactions between individual atoms (or local chiral
centers) on the two molecules.  Contributions to
$\kappa_{IJ}$ when one molecule is excited in the virtual state are of order
$R_{IJ}^{-7}$ for helical molecules which are assumed not to
have a global dipole moment, but whose atoms posses a
dipole moment. It is shown that for a helical molecule $Q$ can have
either the same or the opposite sign as the chiral pitch of an
individual molecule, depending on the details of the anisotropy of the
atomic polarizability.  The one--molecule mechanism can become important
when the local atomic dipoles become sizable, although biaxial
correlations (ignored here) should then be taken into account.
Our results suggest how the architecture of molecular dipole moments
might be adjusted to significantly influence the macroscopic pitch.
\end{abstract}

\newpage
\section{Introduction.}

In the cholesteric liquid crystalline phase,\cite{DEG,TSYK}
anisotropic mesogens align on average along a local unit
director ${\bf n} ( {\bf r} )$
that rotates in a helical fashion about a uniform pitch axis.
The pitch $P$ of this helix ranges from a few tenths of a micron to
ten or more microns.  In fact, solutions of the viruses FD and
TMV, as well as DNA, have even much larger pitches.\cite{FRADEN,PODG}
Because the pitch is usually large compared
to the intermolecular separation, these systems are locally
essentially indistinguishable from nematics and consequently they
are often referred to as chiral nematics (CN's).
The pitch wavenumber $Q =  2 \pi/P$ can
even pass through zero as a function of temperature.\cite{DEG,TSYK}
The helical structure of a cholesteric phase must result from
the molecular chirality of some or all of its constituent mesogens.
Achiral mesogens form an achiral nematic rather than a chiral
nematic phase.  Phenomenologically, the explanation of the twist
of the cholesteric phase is straightforward: chiral mesogens
must lead to a chiral term $h {\bf n} \cdot \nabla \times {\bf n}$
in the long--wavelength free energy density that favors twist.
This tendency to twist is resisted by a twist elastic energy density
$\case 1/2 K_2 ({\bf n} \cdot \nabla \times {\bf n})^2$,
where $K_2$ is the Frank elastic constant for twist.
If the pitch axis coincides with the $z$--direction, then in the
equilibrium configuration one has
\begin{equation}
\label{QEQ}
{\bf n} ( {\bf r} ) = ( \cos Q z, \sin Q z, 0) \ ,
\end{equation}
with
\begin{equation}
\label{EQEQ}
Q \equiv 2 \pi / P = h / K_2 \ .
\end{equation}
(Our definition of $Q$ is such that positive $Q$ corresponds
to right--handed macroscopic chirality.\cite{SIGN}) The
magnitude of $K_2$, which has units of energy per unit length,
is estimated with good accuracy by dimensional analysis.
The characteristic energy is of order the thermal energy,
$k_B T \sim k_B T_{NI}$ where $k_B$ is the Boltzman constant,
$T$ is the temperature, and $T_{NI}$ is the isotropic--to--nematic
transition temperature.  The characteristic length is a molecular
length $L$, so that $K_2 \sim k_B T / L$.
A similar dimensional analysis for the torque field $h$, which
has units of energy/(length$^2$), would predict $h \sim k_BT/L^2$
and $P \sim L$.  This is a far tighter pitch than is observed
in any cholesteric.  This reasoning indicates that an explanation
of the magnitude of $h$ requires considering a
detailed model of the cholesteric.  The chiral structure of
cholesterics also raises some technological issues.
It would be very desirable to be able to ``engineer" molecules
that have specific values of $h$ and thus $P$ or more generally
that have a specific temperature dependence for $h$.
To realize this goal, it is necessary to understand how
variations in molecular architecture and electronic structure
influence $h$.  As a first step in dealing with these issues of
fundamental and applied science, this paper will address some
aspects of the calculation of $h$ from a molecular model.

In fact, the calculation of $h$ is highly nontrivial.  It involves the
rather complex interactions between mesogens and the orientational
correlations they induce.  If there are chiral mesogens,
there are chiral interactions, and $h$ is nonzero;
otherwise, $h$ is zero.  One typically
identifies three types of interactions between molecules:\cite{CRAIG}
(1) long--range attractive dispersion (Van der Waals) interactions,
(2) short--range repulsive interactions, whose origin is
the Pauli principle, and
(3) direct Coulomb interactions, which take the form of dipolar,
quadrupolar, etc. interactions between electrically neutral
mesogens.  The latter interactions are of secondary importance
in many chiral and achiral liquid
crystals and will be ignored here.  Initially Straley\cite{Straley}
proposed that the macroscopic chirality of CN's could be understood
qualitatively in terms of the packing of screws.
These short--range repulsive forces were modeled as hard--core or
steric potentials,\cite{ONSAGER,FRENKEL,EVANS,ITALY} reflecting
molecular shape, that contribute to the entropy but not the internal
energy.  For spherical atoms, the repulsive and dispersion forces
can be combined in a single effective potential such as the
Lennard-Jones $6-12$ central--force potential.  More generally,
interactions between
achiral molecules can be modeled as sums over central--force effective
potentials between pairs of atoms or mass points on different
molecules.\cite{MOLPOT}  There are chiral versions of both
dispersion and short--range repulsive forces.  Chiral dispersion
forces were first analyzed by Goossens\cite{GOOS} and later
more systematically by others.\cite{CRAIG,Meer,Kats}
They found that the dominant chiral interaction between
chiral mesogens, calculated in the limit of center--of--mass
separation $R$ much larger than any molecular dimension
$L$ was proportional to $R^{-7}$ and to the product of dipolar
and quadrupolar molecular matrix elements.  Various somewhat
ad hoc chiral intermolecular interactions, some based on implementing
models equivalent to threaded rods,\cite{BERNE,TSYKALO,SCHRODER,PELC}
others on surface--nematic interactions of chiral dopants,\cite{ITALY}
have been introduced mostly as input to simulations of chiral systems.
Models for flexible mesogens have also been treated.\cite{ODIJK}

A chiral molecule is one that cannot be rotated into coincidence
with its mirror image.\cite{THOMSON}
Chiral molecules cannot be uniaxial: at minimum,
their description requires an orthonormal triad of vectors rather than a
single vector.  A microscopic description of chiral interactions involves
the complete orthonormal triad of axes emblazoned on each of the two
interacting molecules.  However, as we have mentioned, apart from very
small corrections arising from slow local twist, the cholesteric phase
is locally uniaxial.  It is, therefore, natural to seek effective
chiral interactions between effectively uniaxial molecules.
If a molecule $J$ of arbitrary shape is spun about some axis ${\bf a}_J$,
it becomes on average uniaxial with respect to this axis.  Thus, general
pair interactions between molecules $I$ and $J$ in a chiral nematic
can be reduced to uniaxial pair interactions by averaging over
independent rotations of each member of the pair about the local
nematic director.  The resulting potential is only approximate in
that it ignores orientational correlations between molecules in the
plane perpendicular to ${\bf a}_I$ and ${\bf a}_J$.  In practice,
one usually averages over independent rotations of each molecule
about its body axis ${\bf a}$, rather than the more correct average
over rotations about the local nematic director.
We mention that it is known that the chiral part of central--force
potentials (such as hard--body interactions)
vanishes when such correlations are neglected.\cite{SALEM,HKL}
However, dispersion contributions to the chiral
interaction do not require nonzero
orientational correlations between molecules.
As discussed in Appendix \ref{BLURB} and as has been found by
several previous authors,\cite{Meer,Kats}
the long--range dispersion interaction
survives this independent rotation procedure to produce an effective
chiral potential between effectively uniaxial mesogens of the form
\begin{equation}
\label{EIJEQ}
E_{IJ} ( {\bf a}_I, {\bf a}_J, {\bf R}_{IJ} ) =
( {\bf a}_I \times {\bf a}_J \cdot {\bf R}_{IJ} ) \kappa_{IJ} ,
\end{equation}
where ${\bf R}_{IJ}$ is the displacement of the
center of molecule $I$ relative to the center of molecule $J$
and only terms in $\kappa_{IJ}$ which are odd in both ${\bf a}_I$ and
${\bf a}_J$ are retained.
The effective interaction of Eq. (\ref{EIJEQ}) arises
between two chiral molecules as well as between
a chiral molecule and an achiral one.

Our derivation of the effective chiral interaction
differs from previous ones\cite{Meer,Kats} in
two important respects.  First, previous calculations of
this interaction are based on a multipole expansion in the
variable ${\bf r}_i/R_{IJ}$, where ${\bf r}_i$ is the
coordinate of the $i$th charge of molecule $I$ relative
to the center of molecule $I$.  Strictly speaking, the
multipole expansion
only applies when ${\bf R}_{IJ}\equiv {\bf R}$ is large
compared to any dimension of the molecules.  This expansion
does not apply to pairs of molecules whose separation is
less than their length but greater than their width.
We develop a modified multipole expansion in which
coordinates transverse to long molecular axes are treated
as small parameters.  Second, the results of previous
calculations are expressed in terms of electric dipole and
quadrupole matrix elements of the entire molecule.  
But in a long molecule, typical of those comprising liquid
crystals, we expect the electronic states to be strongly
localized.\cite{LOCALIZED}  Accordingly, it seems more useful to
express results in terms of matrix elements within atoms or
local complexes.  In so doing, it is natural to assume the relevant
excited state can be reached from the ground state by matrix
elements of the dipole moment operator.  Then, the quadrupole 
moment operator
is easily related to the dipole moment operator, with the result
that the only matrix elements appearing in the present paper
are those of the dipole moment operator between local atomic states.
In common with previous treatments, we will neglect the effects
of biaxial correlations between interacting molecules.
Accordingly, we will evaluate $\kappa_{IJ}$ by averaging
each molecule independently over spinning about its long axis.
In a separate paper\cite{IH} we will discuss how the chiral
interaction between helical molecules depends on the angles
describing rotation about their longest body axis. 

We may summarize briefly the results of this program.  Although
we do not expand in powers of the longitudinal coordinates
of the charges in each molecule, our results are formally not
very different from the previous ones.\cite{Meer,Kats}  However,
by expressing the results in terms of matrix elements of
localized atomic orbitals, we identify two distinct physical
mechanisms. The first is the dipole--quadrupole interaction previously
identified. The second is one involving a three--body interaction
between two local atomic dipole moments on one molecule and a local
anisotropic polarizability of the second molecule.  This second
interaction, formally present in previous work, can dominate the
first one in certain situations.  Furthermore, our approach allows us
to discuss how these interactions depend on the length of the molecule.
For a helical molecule we find that the contribution to $\kappa_{IJ}$
due to the first mechanism is proportional to $L^2/R^8$ for $L \ll R$ and
to $L/R^7$ for $L \ge R$.  Results for the three--body interaction are
more complicated [see Eq. (\ref{KAPRESEQ}), below].
In both cases, the magnitude of the pitch arising
from these interactions in a concentrated system of helical
molecules with polarizability corresponding to a dielectric
constant of about 1.3 would be 10 microns.  This is a larger pitch
than one observes for most concentrated cholesterics.  It is
possible that this discrepancy is due to some of the simplifying
assumptions in our calculations most probably our disregard of
biaxial correlations between molecules.\cite{IH}
Alternatively, it is possible
that the pitch of most cholesterics is determined by steric rather
than by quantum interactions.
Elsewhere we will apply the approach of the present paper
to obtain the quantum contributions to $K_2$.\cite{IH}

Briefly, this paper is organized as follows.
In Sec. II we give an overview of the calculation of the
torque field $h$ from which the macroscopic chiral pitch can
be determined.  In Sec. III we derive a rather general
expression for the strength of the chiral interaction
between molecules in terms of matrix elements of the
dipole moment operators of atoms between the ground state
and excited states localized on atoms.  The contribution to
this effective chiral interaction from  virtual states in which
both molecules are excited is treated in Sec. IV and that in
which only one molecule is excited is treated in Sec. V.
Numerical estimates of the pitch for a system of helical molecules
are given which show that the quantum mechanism when both molecules 
are excited is unlikely to explain the observed pitch of most
cholesterics.  Our results and conclusions are summarized in Sec. VI.

%%%%%%%%%%%%%%%%%%%%%%%%%%%%%%%%%%%%%%%%%%%%%%%%%%%%%%%%%%%%%%%%%%%%%%%%
%         SECTION 2
%%%%%%%%%%%%%%%%%%%%%%%%%%%%%%%%%%%%%%%%%%%%%%%%%%%%%%%%%%%%%%%%%%%%%%%%

\section{OVERVIEW OF CALCULATION}

In this section we give an overview and a summary of the results of the
calculation, given in the next section, of the chiral interaction between
molecules.  Recently\cite{HKL} a systematic formulation was given that
expresses the macroscopic pitch of a CN in terms of microscopic
interactions between molecules.  Such a formulation is required in cases
where it is either necessary or desirable to include orientational
correlations between interacting molecules.  It was shown that
for central--force interactions between atoms on different molecules,
a nonzero effective chiral interaction between molecules could only
be obtained when orientational correlations, specifically biaxial
correlations, between molecules were taken into account. In contrast,
in the present paper we will see that the quantum interactions
between molecules are not of this type.  Thus, in the present
context, it is permissible to use a simpler and
more traditional approach in which each molecule is characterized
by the orientation of its long axis, specified by the unit vector
${\bf a}$.  We will then evaluate the chiral interaction energy
between molecules $I$ and $J$ written in Eq. (\ref{EIJEQ}).
This interaction energy is evaluated within what
we will call the ``uniaxial" approximation in which we independently
average over the orientations of the two molecules when ${\bf a}$ is
specified for each molecule.  In order to calculate the macroscopic
chiral pitch it is only necessary to evaluate the chiral part of this
interaction, i. e. the part of the form written in Eq. (\ref{EIJEQ}).
This interaction energy can then be added to whatever phenomenological
interaction one is using to describe the nematic phase which would
result in the absence of chiral interactions.

To make contact with a continuum theory, one introduces a
local order parameter tensor via
\begin{eqnarray}
Q_{\alpha \beta } ({\bf r}) = a_\alpha ({\bf r})
a_\beta ({\bf r}) - \case 1/3 \delta_{\alpha \beta} \ .
\end{eqnarray}
When thermally averaged, this tensor becomes the usual
de Gennes--Maier--Saupe order parameter.\cite{Maier,DEG}
In the long--wavelength limit, the chiral interaction $E_{IJ}$
leads to a continuum interaction of the form
\begin{eqnarray}
E_{\rm int} = \gamma' \int d {\bf r} \epsilon_{\alpha \beta \gamma}
Q_{\alpha \delta}({\bf r}) \nabla_\beta Q_{\gamma \delta}({\bf r}) \ ,
\end{eqnarray}
where Greek indices label Cartesian components, $\delta_{\alpha \beta}$
is the Kronecker delta, $\epsilon_{\alpha \beta \gamma}$ is the
antisymmetric tensor, the repeated--index summation convention is
understood,
and the constant $\gamma'$ is the macroscopic analog of $\kappa_{IJ}$.
If we express ${\bf Q}$ in terms of the director ${\bf n}({\bf r})$ as
\begin{eqnarray}
\langle Q_{\alpha \beta}\rangle_T
 = S \left( n_\alpha n_\beta - \case 1/3 \delta_{\alpha \beta} \right) \ ,
\end{eqnarray}
where $\langle \ \ \rangle_T$ indicates a thermal average,
then the above chiral interaction leads to the familiar
Frank free energy in the presence of macroscopic chiral twist (but
neglecting splay and bend distortions) as\cite{SIGN}
\begin{eqnarray}
\label{FRANK}
F &=&  \case 1/2 K_2 \int d {\bf r} [
{\bf n}({\bf r}) \cdot \nabla \times {\bf n}({\bf r})]^2 +
h \int d{\bf r}  {\bf n}({\bf r}) \cdot \nabla \times {\bf n}({\bf r}) \ ,
\end{eqnarray}
where, within
mean--field theory, $h=\gamma' S^2$.  The main goal of the present paper
is to obtain $h$ from a microscopic model.  In particular we will obtain
$h$ from an evaluation of $\kappa_{IJ}$ in Eq. (\ref{EIJEQ}).
As we have seen in Eqs.  (\ref{QEQ}) and (\ref{EQEQ}) a determination of
$h$ leads immediately to the determination of the macroscopic chiral wave
vector ${\bf Q}$.  In addition, the
chiral properties of an isotropic liquid consisting of chiral
molecules, such as the rotary power, are also related to $\kappa_{IJ}$.

\begin{figure}
\psfig{figure=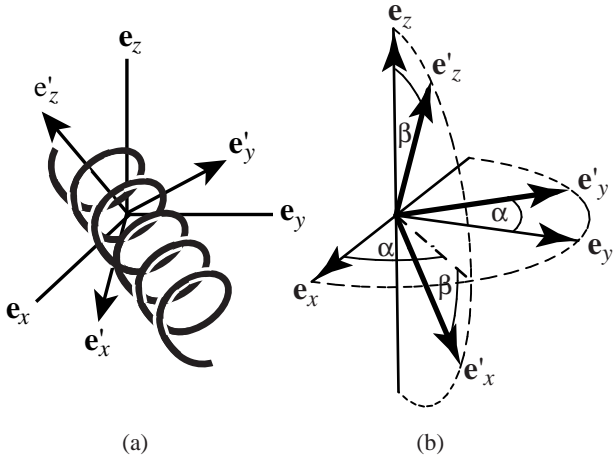}

\vspace{0.2 in} \noindent
\caption{Left: Molecule--fixed coordinate system, defined by the unit
vectors ${\bf e}'_\mu$.  Right: Definition of the Euler angles $\alpha$,
$\beta$, and $\gamma$ which take the space--fixed axes ${\bf e}_x$,
${\bf e}_y$, and ${\bf e}_z$ into the molecule--fixed axes,
${\bf e}'_x$, ${\bf e}'_y$, and ${\bf e}'_z$.  Note that $\alpha$
and $\beta$ are the usual spherical angles which specify the
orientation of the long axis of the molecule, ${\bf e'}_z$, with
respect to the space--fixed axes.  The third Euler angle $\gamma$,
not shown here, is the angle of rotation about the $z'$ axis which
brings the $x$ and $y$ axes in coincidence with the
$x$ and $y$ axes fixed in the body (respectively ${\bf e}'_x$ and
${\bf e}'_y$).}
\label{EULER}
\end{figure}

\begin{figure}
\psfig{figure=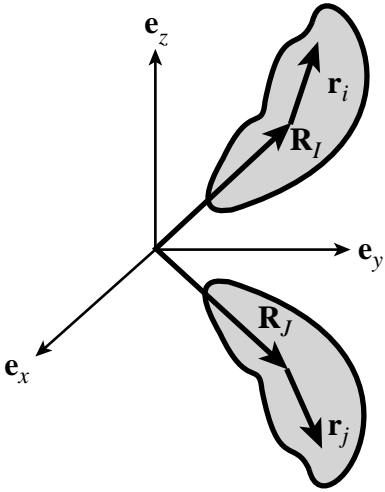}

\vspace{0.2 in} \noindent
\caption{Space--fixed coordinate system, showing the displacement,
${\bf R}_I$, of the $I$th molecule and the displacement, ${\bf r}_i$
of the $i$th
charge of the $I$ molecule relative to the center of the molecule.}
\label{SPACEF}
\end{figure}

The interaction between molecules we are going to study is the
generalization, for chiral molecules, of the attractive $1/R^6$
term in the van der Waals potential between neutral spherical atoms.
This calculation is based on a quantum mechanical treatment of the
total Coulomb interaction ${\cal H}_{IJ}$ between charges on the two
interacting molecules, $I$ and $J$:
\begin{eqnarray}
\label{VEQ}
{\cal H}_{IJ} &=& \sum_{i \in I} \sum_{j \in J} {q_i q_j \over R_{ij} } \ ,
\end{eqnarray}
where $i \in I$ indicates that the sum is over all charges $q_i$, both
electronic and nuclear, in molecule $I$ and $R_{ij}$ is the displacement
of $q_i$ relative to $q_j$.
In this calculation we neglect any biaxial correlations between the
orientations of the two molecules.  Within this assumption it has
been shown\cite{HKL} that central force interactions [like those
of Eq. (\ref{VEQ}) when taken in first order perturbation theory]
can not lead to any chiral interactions.  Therefore, we consider
the effect of ${\cal H}_{IJ}$ within second--order perturbation theory.
The effective interaction between molecules $I$ and $J$ is then
\begin{eqnarray}
E_{IJ} = E_{IJ}( \hat \omega_I , \hat \omega_J , {\bf R}_{IJ}) =
\langle 0 | {\cal H}_{IJ} {{\cal P} \over {\cal E}} {\cal H}_{IJ}
|0 \rangle \ ,
\end{eqnarray}
where ${\cal E} = E_0 - {\cal H}_0$, with $E_0$ the energy of
the ground state $|0 \rangle$ of ${\cal H}_0$ the unperturbed
Hamiltonian describing noninteracting molecules,
and ${\cal P}$ is a projection operator that excludes
the ground state.  Here we have indicated that $E_{IJ}$ depends on
$\hat \omega_I$ (and $\hat \omega_J$) which denotes the triad of
Euler angles (see Fig. \ref{EULER}) needed to specify the orientation
of the $I$th ($J$th) molecule and on ${\bf R}_{IJ}$.  Both
${\bf R}_{IJ}$ and ${\bf r}_i$, the position of the $i$th charge
in molecule $I$ relative to the center of the molecule, ${\bf r}_i$,
may be expressed with respect to coordinate axes, ${\bf e}_\mu$,
fixed in space, as shown in Fig. \ref{SPACEF}:
\begin{eqnarray}
{\bf R}_I &=&  R_{I\mu} {\bf e}_\mu \ , \ \ \ 
{\bf r}_i =  r_{i\mu} {\bf e}_\mu \ .
\end{eqnarray}

We now must average this interaction energy over the orientations of the
two molecules, when the long axes of molecules $I$ and $J$ are
fixed to lie along the unit vectors ${\bf a}_I$ and ${\bf a}_J$,
respectively, and correlations between the orientations of the two
molecules are neglected.  To carry out this average we introduce axes
specified by unit vectors ${\bf e}'_{I\mu}$ emblazoned on the
$I$th molecule, as shown in Fig. \ref{EULER}, so that
\begin{eqnarray}
\label{EMBLAZE}
{\bf r}_i &=& r'_{i \mu} {\bf e}'_{I\mu} \ .
\end{eqnarray}
Previously van der Meer et al.\cite{Meer} carried out this averaging
within the multipole expansion.  However, since we wish to treat
long molecules of the type usually constituting liquid crystals,
we do not make the usual multipole expansion, but rather expand
only in terms of transverse coordinates of the molecule.  Thus we set
\begin{eqnarray}
\label{RHOEQ}
{\bf r}_i= z'_i {\bf a}_I + \rhov_i \ ,
\end{eqnarray}
where $\rhov_i \cdot {\bf a}_I =0$.  [Throughout, atom $i$
($j$) is assumed to be in molecule $I$ ($J$)].
Thus $z'_i$ and $\rhov _i$ are the coordinates
of the $i$th charge of molecule $I$ relative to the center of the
molecule, respectively, longitudinal and transverse to the long axis
of the molecule aligned along ${\bf a}_I = {\bf e}'_z$.  Now we
expand $E_{IJ}$ in powers of $\rhov$ and perform the orientational
average over powers of $\rhov$ (indicated by $[\ \ ]_{\rm av}$)
is done using, e. g.,
\begin{eqnarray}
\label{AVEEQ}
{ [ (\rho_i)_\alpha (\rho_{i'})_\beta ]_{\rm av} } & = &
\case 1/2 (x'_i x'_{i'} + y'_i y'_{i'})(\delta_{\alpha , \beta}
- a_\alpha a_\beta ) + \case 1/2
\epsilon_{\alpha \beta \gamma} a_\gamma (x'_i y'_{i'}-y'_i x'_{i'})
\nonumber \\
&=& \case 1/2  r'_{i\mu} r'_{i'\mu} (\delta_{\alpha,\beta}
-a_\alpha a_\beta ) + \case 1/2 \epsilon_{\mu \nu z} r'_{i\mu} r'_{i'\nu} 
\epsilon_{\alpha \beta \gamma} a_\gamma \ ,
\end{eqnarray}
where $\mu$ and $\nu$ run over only transverse ($x$, $y$) coordinates
and $a_\gamma \equiv ({\bf a}_I)_\gamma$.  Thereby we find that
\begin{eqnarray}
[ E_{IJ} ]_{\rm av} &=& ( {\bf a}_I \times {\bf a}_J \cdot {\bf R}_{IJ} )
\kappa_{IJ} ( {\bf a}_I , {\bf a}_J, {\bf R}_{IJ} )  + \dots \ .
\end{eqnarray}
Here we have written the term responsible for the chiral interaction
between molecules $I$ and $J$ and have discarded the nonchiral terms
(represented by $\dots$).

The expressions for $\kappa_{IJ}$ in its most general form are
not very enlightening, although they do display the appropriate
symmetry to vanish for molecules which are not chiral.  To gain
some insight into the meaning of these results we have had recourse
to a model of the excited states, which appear in second--order
perturbation theory.  Our first assumption is that the important
excited states consist of dipolar fluctuations from the ground state.
In other words these states are taken to be the three atomic $p$
states $|\mu_i \rangle$ on atom $i$.  The second assumption is that
$\delta$, the width in energy
of the band of excited states obtained by allowing these excitations to
occur on any atom is small compared to their energy $E$ relative
to the ground state.  This assumption allows us to take the
virtual intermediate states to be strictly localized to individual
atoms.\cite{LOCALIZED}  Nonlocal effects give rise to corrections of
relative order, $\delta / E$.  Under these assumptions, our results
may be summarized as follows.  Contributions to $\kappa_{IJ}$
can be classified into two types, depending on whether one or both
molecules in the intermediate state are in an excited state.  These
are denoted $\kappa^{(1)}_{IJ}$ and $\kappa^{(2)}_{IJ}$, and will be
referred to as ``one--molecule" and ``two--molecule" terms, respectively.
Our results are conveniently written in terms of the definition
$\kappa^{(n)}_{IJ}= \case 1/2 [ \tilde \kappa^{(n)}_{IJ}
+ \tilde \kappa^{(n)}_{JI}]_{\bf a}$,
where, for any function $f$ of ${\bf a}_I$ and ${\bf a}_J$,
\begin{eqnarray}
\label{AAVE}
\left[ f \right]_{\bf a}  = \case 1/4 [ f({\bf a}_I, {\bf a}_J)
- f(-{\bf a}_I, {\bf a}_J) - f({\bf a}_I, -{\bf a}_J)
+ f(-{\bf a}_I, -{\bf a}_J) ]\ .
\end{eqnarray}
Then
\begin{eqnarray}
\label{K2RESULT}
\tilde \kappa_{IJ}^{(2)} &=& \sum_{i \in I, j \in J } M_{ij} S_{ij} \ ,
\end{eqnarray}
where
\begin{eqnarray}
S_{ij} = [{\bf a}_I \cdot {\bf a}_J
-2({\bf a}_I \cdot \overline {\bf D}_{ij})
({\bf a}_J \cdot \overline {\bf D}_{ij} )/{\overline D}_{ij}^2 ]
\overline D_{ij}^{-8}
\end{eqnarray}
\begin{eqnarray}
\label{K2FORMULA}
M_{ij} & = & 3e^4 \Biggl\{ \sum_{\mu ,\nu } E_{\nu \mu}(i,j)^{-1}
\Biggl[ \overline y'_j \langle 0 | \Delta z'_j | \mu_j \rangle
\langle \mu_j | \Delta x'_j | 0 \rangle
- \overline x'_j \langle 0 | \Delta z'_j |\mu_j \rangle
\langle \mu_j | \Delta y'_j | 0 \rangle \Biggr] \nonumber \\ && \
\times \Biggl[ 2 \langle 0 |\Delta z'_i|\nu_i \rangle^2
- \langle 0 |\Delta x'_i| \nu_i \rangle^2
- \langle 0 |\Delta y'_i| \nu_i \rangle^2 \Biggr] \Biggr\} \ ,
\end{eqnarray}
where the sums over $i$ and $j$ now run only over electrons,
$\overline {\bf r}'_j$ is the expectation value of ${\bf r}'_j$
in the ground state, i. e.  it is the center of the atom associated with
charge $j$, $\Delta {\bf r}'_j = {\bf r}'_j = \overline {\bf r}'_j$ and
\begin{eqnarray}
\label{DBAR}
\overline {\bf D}_{ij} = {\bf R}_{IJ}
+ {\overline z}'_i {\bf a}_I - {\overline z}'_j {\bf a}_J \ .
\end{eqnarray}
Also $|\mu_j\rangle$ denotes the state when all atoms are in
their ground state except for atom $j$, which is in the excited
$p$ state labeled $\mu$, which has energy $E_\mu(j)$ relative to
the ground state and $E_{\nu \mu}(i,j)=E_\nu(i)+E_\mu(j)$.  Here
$\langle \mu_j | \Delta {\bf r}_j | 0 \rangle$ is nonzero
only when $j$ refers to an electronic charge.  We also find that
\begin{eqnarray}
\label{K1RESULT}
\tilde \kappa_{IJ}^{(1)} & = & 3 \sum_{ii' j}
[p'_{ix} p'_{i'y} - p'_{i'x} p'_{iy} ]
\frac{(\overline {\bf D}_{i'j} \cdot {\bf a}_J)}
{\overline D_{ij}^3 \overline D_{i'j}^5} \nonumber\\
&\times & e^2 \sum_{\mu } E_\mu(j)^{-1} 
[2\langle \mu_j | \Delta z'_j |0 \rangle^2
-  \langle \mu_j | \Delta x'_j |0 \rangle ^2
-  \langle \mu_j | \Delta y'_j |0 \rangle ^2 ] \ ,
\end{eqnarray}
where the sums over $i$ and $i'$ are over atoms and ${\bf p}'_\alpha$
is the $\alpha$ component of the dipole moment vector in the
ground state evaluated in the molecule--fixed coordinate system.
(For this calculation a local dipole moment was assumed,
but this does not necessarily imply the existence of a dipole
moment of the molecule as a whole.)  Note that the above expressions,
since they have already been averaged over rotations about the long
axis, are invariant with respect to rotation of each molecule
about its long axes parallel to ${\bf e}'_z$.

Our result for $\kappa^{(2)}_{IJ}$ is closely related to that previously
obtained by van der Meer et al\cite{Meer} and by Kats.\cite{Kats} To
obtain a form close to that obtained by Kats, we write
\begin{eqnarray}
M_{ij} & = & {1 \over 2 \pi i} \int_{-\infty}^\infty
\sigma' (j;\omega+i0^+) \gamma' (i;- \omega+i0^+) d \omega \ ,
\end{eqnarray}
where
\begin{eqnarray}
\sigma' (j; \omega ) & =& e^2 \sum_\mu \left(
\langle 0 | y'_j z'_j | \mu_j \rangle \langle \mu_j | x'_j | 0 \rangle 
- \langle 0 | x'_j z'_j | \mu_j \rangle \langle \mu_j | y'_j | 0 \rangle 
\right) / [\omega - E_\mu(j) ]
\end{eqnarray}
\begin{eqnarray}
\gamma' (i; \omega) & =& e^2 \sum_\nu
\left( 2\langle 0 | z'_i | \nu_j \rangle^2
-\langle 0 | x'_i | \nu_j \rangle^2
-\langle 0 | y'_i | \nu_j \rangle^2 \right)
/ [\omega - E_\nu(i) ] \nonumber \\
& = & \alpha'_{zz}(i;\omega) - \case 1/2 \alpha'_{xx}(i;\omega)
- \case 1/2 \alpha'_{yy}(i,\omega) \ ,
\end{eqnarray}
where $\alpha'_{\mu \nu}$ is the $\mu$-$\nu$ component of the 
polarizability tensor with respect to the molecular frame.
Here $\gamma'(\omega)$ is the anisotropy of the polarizability
and $\sigma'(\omega)$ is the higher order quadrupole-dipole response
function (which Kats calls the gyrotropy), both taken in the
molecular frame, as indicated by prime superscripts.  Here these
quantities are given by a sum over the corresponding properties
for the individual atoms.  We assume that the relevant excited
states are localized $p$ states, in which case the gyrotropy can
be related to the polarizability:
\begin{eqnarray}
\label{SIGMAEQ}
\sigma' (j; \omega ) & =& e^2 \sum_\mu \left(
\overline y'_j \langle 0 | z'_j | \mu_j \rangle
\langle \mu_j | x'_j | 0 \rangle 
- \overline x'_j \langle 0 | z'_j | \mu_j \rangle
\langle \mu_j | y'_j | 0 \rangle  \right)
/ [\omega - E_\mu(j) ] \nonumber \\
& = & \case 1/2 \overline y'_j  \alpha_{zx}(j;\omega) -
\case 1/2 \overline x'_j \alpha_{yz}(j;\omega) \ .
\end{eqnarray}
In addition the factor $S_{ij}$ depends on the $z'$-component
of the position of the $i$th atom, whereas in the bare multipole
expansion used by van der Meer and Kats, only ${\bf R}_{IJ}$ appears.
Because we do not include the $z'$ coordinate within the multipole
approximation, we can treat long molecules in an appropriate way,
as is reflected in the sum over atoms of $S_{ij}$.
One sees that the chirality of the molecule is incorporated in
$\sigma'$, which vanishes if the molecule has a mirror
plane.\cite{MIRROR} In $\tilde \kappa^{(1)}_{IJ}$, the chirality
of the molecule $I$ is incorporated in terms like
\begin{eqnarray}
\tau' \equiv \sum_{i i'} [p'_{ix} p'_{i'y} - p'_{i'x} p'_{iy}]
[\overline z'_i - \overline z'_{i'}] \ .
\end{eqnarray}
In the case of classical interactions, it was not possible to
construct a third rank tensor of the mass moments which was zero
for achiral molecules and nonzero for chiral molecules,\cite{HKL}
because such a mass moment tensor was symmetric under
interchange of any pair of its three indices.  Here, however,
one sees that $\sigma'$ and $\tau'$ are $x$,$y$,$z$
elements of tensors that are not symmetric in all indices
and that, therefore, can be used as an indicator of chirality.

\begin{figure}
\psfig{figure=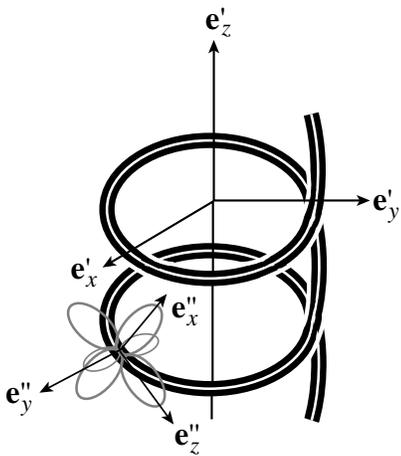}

\vspace{0.2 in} \noindent
\caption{Local atomic coordinate system, defined by the unit vectors
${\bf e}''_\mu$, showing that the local excited $p$ states define
the orientation of the local axes.  Here ${\bf e}''_z$ is the unit
vector tangent to the helix, the unit normal, ${\bf e}''_x$, lies along
the radius of curvature, and the binormal unit vector ${\bf e}''_y$
is the third member of the triad of mutually perpendicular unit vectors.}
\label{ATOMF}
\end{figure}

It is interesting to evaluate these expressions for some specific
geometry of a chiral molecule.  For this purpose we treat in some
detail a helical molecule, patterned after DNA.  Then we introduce
local atomic coordinates whose axes coincide with the axes defined
by the local excited $p$ states.  We assume that these axes, shown in
Fig. \ref{ATOMF},  are identical to those of the tangent, the normal,
and the binormal unit vectors, which we call ${\bf e}''_\mu$, with $\mu$
respectively, $z$, $x$, and $y$, so that we can relate the $\alpha'_{\mu,\nu}$
to its components $\alpha''_{\mu,\mu}$ in the local atomic frame as
\begin{eqnarray}
\label{LOCALTR}
\alpha'_{\gamma \delta}(j) = 2 e^2 \sum_{\mu} E_{\mu}(j)^{-1}
\langle \mu_j | r''_{j \mu} | 0 \rangle^2
({\bf e}''_{j\mu} \cdot {\bf e}'_\gamma)
({\bf e}''_{j\mu} \cdot {\bf e}'_\delta) =
({\bf e}''_{j\mu} \cdot {\bf e}'_\gamma)
({\bf e}''_{j\mu} \cdot {\bf e}'_\delta) \alpha''_{\mu \mu}(j) \ .
\end{eqnarray}

One should note the following general points in connection with
our results.  Firstly, the result in Eq. (\ref{K2RESULT}) shows that
$\kappa^{(2)}_{IJ}$ can be viewed as arising from a superposition
of interactions between local centers of chirality on one
molecule with centers of anisotropic polarizability on another
molecule.  As is well known,\cite{Meer} this result implies
that chirality can be induced by the interaction between a chiral
molecule and an achiral one that has a local center of anisotropic
polarizability.  In contrast, the result in Eq. (\ref{K1RESULT})
is a three--body interaction between two local dipoles
on one molecule (combined with resulting chiral strength $\tau'$)
with a local anisotropic polarizability of the second molecule.
Finally, we mention that it is interesting to generalize these
results to a flexible polymer the orientation of whose backbone
may vary appreciably over its length.

\section{CHIRALITY FROM INTERMOLECULAR INTERACTIONS}

We now turn to the calculation of $\kappa_{IJ}$.  For this purpose
we give a brief discussion of how the average over orientations
is to be done.  In general, the orientation of the $I$th molecule is
specified by the three Euler angles $\alpha_I$, $\beta_I$, and $\gamma_I$,
for which we adopt the definition of Rose,\cite{ROSE} as is illustrated
in Fig. \ref{EULER}.  In particular, $\alpha_I$ and $\beta_I$ are taken
to specify the orientation of the long axis of the molecule.  So we write
\begin{eqnarray}
{\bf a}_I = \sin \beta_I \cos \alpha_I {\bf e}_x + \sin \beta_I
\sin \alpha_I {\bf e}_y + \cos \beta_I {\bf e}_z \ .
\end{eqnarray}
Within the spirit of mean--field theory we should average the
interaction energy between molecules $I$ and $J$ over the
single--molecule orientational distribution function appropriate
to a nematic, which locally is a good description of the CN.
For molecule $I$ this average should be taken subject to its long
axis being specified by the fixed value of ${\bf a}_I$.  The
single molecule orientation distribution function $\rho(\hat \omega_I)$
must be independent of $\alpha_I$ and also should be invariant under
${\bf a} \rightarrow - {\bf a}$.  Specifically, correlations between
$\beta_I$ and $\gamma_I$ are allowed,\cite{MOON} as is discussed in
Appendix \ref{MOON}.  For any
function of molecular orientation $f(\hat \omega)$ this average is
\begin{eqnarray}
[f(\hat \omega)]_{\rm av} = {1 \over 2\pi} \int
\rho( \hat \omega_I ) f( \hat \omega_I ) d \gamma_I d \alpha_I 
{\large /} \int \rho(\hat \omega_I ) d \gamma_I \ . 
\end{eqnarray}
However, when the molecule is not strongly biaxial, or when the
molecule is perfectly aligned along the nematic direction, the
assumption that $\rho(\hat \omega)$ is independent of $\gamma_I$,
as is usually done,\cite{Meer,Kats}
is sufficient.  This approximation, which we call the uniaxial
approximation, will be used in this paper.  In addition, to preserve
invariance under ${\bf a}_I \rightarrow - {\bf a}_I$, we will
also project out of the calculation terms in $\kappa_{IJ}$ that
are even in ${\bf a}_I$ and in ${\bf a}_J$.  This step can
be done at the end of the calculation using Eq. (\ref{AAVE}).

Our calculation of $\kappa_{IJ}$, as previous ones,\cite{Meer,Kats}
is analogous to that of the well--known $R^{-6}$ interactions between
widely separated neutral atoms.  Quantum fluctuations involving
dipole moments in excited states are treated within second--order
perturbation theory.  Short-range quantum repulsion is often treated
in an ad hoc fashion via a classical central--force interaction
between atoms but this effect will not be discussed here.
We take the interaction Hamiltonian, ${\cal H}_{IJ}$, for molecules
$I$ and $J$ to arise from the Coulomb interaction between the $i$th
charge on molecule $I$, denoted $q_i$, and its counterpart on
molecule $J$.  Thus we write
\begin{equation}
\label{EQ1}
{\cal H}_{IJ}=\sum_{i \in I} \sum_{j \in J} {q_i q_j \over
\mid {\bf R}_{IJ} + {\bf r}_i - {\bf r}_j \mid } \ .
\end{equation}
We use Eq. (\ref{RHOEQ}) to write
\begin{equation}
{\cal H}_{IJ}=\sum_{i \in I} \sum_{j \in J} {q_i q_j \over D_{ij}}
\left[ 1+ \frac{2}{D_{ij}^2} \rhov_{ij} \cdot
{\bf D}_{ij} + \frac{1}{D_{ij}^2} \rhov_{ij}^2 \right]^{-1/2},
\end{equation}
where $\rhov_{ij} = \rhov_i - \rhov_j$ and
\begin{eqnarray}
{\bf D}_{ij} & = & {\bf R}_{IJ} + z'_i {\bf a}_I - z'_j {\bf a}_J \ .
\end{eqnarray}
Note that ${\bf D}_{ij}$ is evaluated for $\rhov_i = \rhov_j=0$.

We now expand with respect to transverse coordinates to obtain
\begin{eqnarray}
\label{VIJEQ}
{\cal H}_{IJ}& = &\sum_{i \in I , j \in J} \frac{q_i
q_j}{D_{ij}} \Biggl[ 1- \frac{1}{D_{ij}^2}\rhov_{ij}
\cdot {\bf D}_{ij} - \frac{ \rhov_{ij}^2}
{2D_{ij}^2} + \frac{3[\rhov_{ij} \cdot {\bf D}_{ij}]^2} {2D_{ij}^4}
- \frac{5[\rhov_{ij} \cdot {\bf D}_{ij}]^3}{2D_{ij}^6}
\nonumber \\ &&
+ \frac{3[\rhov_{ij} \cdot {\bf D}_{ij}]} {2D_{ij}^4}
\rhov_{ij}^2 + {3  \rhov_{ij}^4 \over 8 D_{ij}^4 }
- {15 \over 4} { [ \rhov_{ij} \cdot {\bf D}_{ij} ]^2 \over D_{ij}^6 }
\rhov_{ij}^2 + {35 \over 8} {[\rhov_{ij} \cdot{\bf D}_{ij}]^4 \over
D_{ij}^8} + {\rm O}(\frac{1}{D_{ij}^5}) \Biggr] \ .
\end{eqnarray}
Note that this expansion is valid if the charge
distributions of the two molecules do not overlap one another.
Strictly speaking, the validity of our treatment requires
satisfying this condition for all configurations with
nonnegligible weight in the partition function.
 
We now consider an evaluation of the interaction energy between
two molecules treating ${\cal H}_{IJ}$ via perturbation theory.
The first term is the ground--state expectation value of
the Coulomb interaction between atoms on different molecules.
If we neglect biaxial correlations between orientations of adjacent
molecules and we simply average this interaction over the uncorrelated
rotations of the two molecules, subject to their long axes being
fixed, then we find the resulting interaction to be completely
achiral.\cite{HKL} Accordingly, to obtain an effective chiral
interaction from ${\cal H}_{IJ}$ when such biaxial correlations
are neglected, it is necessary to evaluate the energy of
interaction within second--order perturbation theory, whereby
\begin{eqnarray}
\label{E2EQ}
E_{IJ} = - {\sum_{n_I , n_J}}^\prime
{\mid ( {\cal H}_{IJ})_{n_I, n_J ; 0,0}
\mid^2 \over E_{n_I n_J} } \ ,
\end{eqnarray}
where the sums are over states $|n_I\rangle$ ($|n_J\rangle$) of
molecule $I$ ($J$) and the prime indicates exclusion of the term
when both molecules are in their ground state.  Here
$E_{n_I n_J}$ is the energy (relative to the ground state)
of the state when molecules $I$ and
$J$ are in states $|n_I\rangle$ and $|n_J\rangle$, respectively.

The obvious step of substituting the expansion of Eq. (\ref{VIJEQ})
into Eq. (\ref{E2EQ}) leads to rather complicated algebra. 
We now classify terms according to their order in $R_{IJ}^{-1}$.
Since we do not assume the length $L$ of the molecules
to be much less than the separation $R_{IJ}$ between molecules,
in counting powers of $R_{IJ}$ we consider $L/R_{IJ} \sim 1$.
As we shall see, $E_{\rm IJ} \sim R_{IJ}^{-p}$,\cite{SCALE}
where $p=7$ for two--molecule terms and $p=6$ for one--molecule terms.
Accordingly, we drop all contributions which are of order
$R_{IJ}^{-p}$ with $p>7$.  Also, we drop contributions which
are proportional to odd powers of $\rhov$, 
since these will vanish when we average over
rotation about the long axis of the molecules.  Thereby we obtain
\begin{eqnarray}
{[ E_{IJ} ]_{\rm av} } &= & - \left[ {\sum_{n_I,n_J}}'
\frac{\mid ({\cal H}_{IJ})_{n_I n_J;00} \mid^2}{E_{n_I n_J}}
\right]_{\rm av}
= - \sum_{i,i' \in I} \sum_{j,j'\in J} {\sum_{n_I,n_J}}'
\left[ {\cal E}_{ij;i'j';n} \right]_{\rm av}
{q_i q_{i'} q_j q_{j'} \over E_{n_I n_J}} \ ,
\end{eqnarray}
where $n$ is shorthand for $n_I, n_J$ and
\begin{eqnarray}
\label{ETWOEQ}
{\cal E}_{ij;i'j';n}&& = \left[ {1 \over D_{ij}} \right]_{0n}
\left[ {1 \over D_{i'j'}} \right]_{n0} 
- \left[ {\rhov_{ij}^2 \over D_{ij}^3}
\right]_{0n} \left[ {1 \over D_{i'j'} } \right]_{n0}
\nonumber \\ && \ + \left[ { 3 (\rhov_{ij}
\cdot {\bf D}_{ij})^2 \over D_{ij}^5} \right]_{0n}
\left[ {1 \over D_{i'j'}} \right]_{n0}
+ {1 \over 4} \left[ {\rho_{ij}^2 \over D_{ij}^3} \right]_{0n}
\left[ { \rho_{i' j'}^2 \over D^3_{i'j'}} \right]_{n0}
\nonumber \\ &&
+ \left[ {\rhov_{ij} \cdot {\bf D}_{ij} \over D_{ij}^3} \right]_{0n}
\left[ {\rhov_{i'j'} \cdot {\bf D}_{i'j'}\over D_{i'j'}^3} \right]_{n0}
- \frac{3}{2} \left[ { (\rhov_{ij} \cdot {\bf D}_{ij})^2 \over D_{ij}^5}
\right]_{n0} \left[ { \rhov_{i'j'}^2 \over D_{i'j'}^3} \right]_{0n}
\nonumber\\ &&+ \frac{9}{4} \left[ { (\rhov_{ij} \cdot
{\bf D}_{ij})^2 \over D_{ij}^5} \right]_{n0} \left[ { (\rhov_{i'j'} \cdot
{\bf D}_{i'j'})^2 \over D_{i'j'}^5}\right]_{0n} \nonumber\\
&&- 3 \left[ { (\rhov_{ij} \cdot {\bf D}_{ij}) \rhov_{ij}^2 \over D_{ij}^5}
\right]_{n0} \left[ { \rhov_{i'j'} \cdot {\bf D}_{i'j'} \over D_{i'j'}^3}
\right]_{0n}
+ 5 \left[ { \rhov_{ij} \cdot {\bf D}_{ij} \over D_{ij}^3} \right]_{n0}
\left[ {(\rhov_{i'j'} \cdot {\bf D}_{i'j'})^3
\over D_{i'j'}^7 } \right]_{0n} \nonumber \\
&& + \left[ {3 \rhov_{ij}^4 \over 4 D_{ij}^5 }
- {15 \rhov_{ij}^2 \over 2 D_{ij}^7 }
(\rhov_{ij} \cdot {\bf D}_{ij})^2 + {35 (\rhov_{ij}
\cdot {\bf D}_{ij})^4 \over 4 D_{ij}^9}
\right]_{n,0}  \left[ {1 \over D_{i'j'}} \right]_{0,n} \ .
\end{eqnarray}
When one averages over independent rotations of the two
molecules about their long axes, using Eq. (\ref{AVEEQ}),
one sees that the first two lines of Eq.  (\ref{ETWOEQ}) do
not lead to a chiral interaction.

We imagine the virtual states $\{n\}$ in Eq. (\ref{ETWOEQ})
to be a linear combination of excited atomic $p$ states.
Accordingly, all matrix elements can be chosen to be real.
Also, in this model we take no explicit account of exchange
and correlation effects beyond what
is included in self-consistent atomic orbitals.  Thus, it is
permissible to label electrons according to their atomic location.
Then, for the matrix element of an arbitrary function, $f$,
of ${\bf r}_i$ we can write
\begin{eqnarray}
\label{MATELEQ}
\langle n_i | f({\bf r}_i) | 0 \rangle =
\nabla'_\alpha f({\bf r}) \mid_{{\bf r}_i = \overline {\bf r}}
\langle n_i | \Delta r'_{i \alpha} | 0 \rangle 
+ {\rm O} \left( \langle n_i | \Delta {\bf r}'_{i\alpha}
\Delta {\bf r}'_{i\beta} | 0 \rangle \right) \ ,
\end{eqnarray}
where $\Delta {\bf r}_i = {\bf r}_i - \langle 0| {\bf r}_i | 0
\rangle \equiv {\bf r}_i - \overline {\bf r}_i$.
To leading order in $1/R_{IJ}$ we have
\begin{eqnarray}
{ \langle n_i | D_{ij}^{-1} | 0 \rangle } &=&
- \langle n_i | \Delta z'_i | 0 \rangle 
({\bf a}_I \cdot \overline {\bf D}_{ij} ) \overline D_{ij}^{-3} \ ,
\nonumber \\ { \langle n_i n_j | D_{ij}^{-1} | 0 \rangle } &=&
- \langle n_i n_j | \Delta z'_i \Delta z'_j | 0_i 0_j \rangle
[3 ({\bf a}_I \cdot \overline {\bf D}_{ij} )
({\bf a}_J \cdot \overline {\bf D}_{ij}) - {\bf a}_I \cdot {\bf a}_J
\overline D_{ij}^2 ] \overline D_{ij}^{-5} \ ,
\end{eqnarray}
where $| n_i n_j \rangle$ is the state (whose energy relative to
the ground state is $E_{n_i n_j}$) in which atom $i$ is in excited
state $|n_i\rangle$, atom $j$ is in state $|n_j \rangle$, and all
other atoms are in their ground state and $\overline {\bf D}_{ij}$
was defined in Eq. (\ref{DBAR}).
Thus $\left[ D_{ij}^{-1} \right]_{0,n}$ is of order at least
$R_{IJ}^{-2}$ for single--molecule terms and of order at least
$R_{IJ}^{-3}$ for two--molecule terms.  This argument shows that
the last line of Eq. (\ref{ETWOEQ}) does not contribute at leading
order in $1/R_{IJ}$ and that we have only to deal with
lines 3, 4, and 5 of this equation.  

We now carry out the average over the orientations of molecules $I$
and $J$ subject to their long axes being fixed to be, respectively,
along ${\bf a}_I$ and ${\bf a}_J$, using Eq. (\ref{AVEEQ}). and we
keep only chiral terms of the form written in Eq. (\ref{EIJEQ}).
This procedure is algebraically extremely complicated.  However,
the fact that the relevant excited states are undoubtedly strongly
localized leads to drastic simplifications.
Accordingly, we will evaluate Eq. (\ref{ETWOEQ}) within a model in
which each molecule has a narrow band of excited states.  If we set
$\kappa_{IJ}= \case 1/2 [\tilde \kappa_{IJ} + \tilde \kappa_{JI}]$,
then the chiral terms which arise from performing the orientational
average (see Appendix \ref{BCAVE}) lead to the result
\begin{eqnarray}
\label{KAPPAEQ}
{\tilde \kappa_{IJ}} &=& \sum_{i,i',j,j'} {\sum_n}'
\frac{q_i q_j q_{i'} q_{j'}}{E_{n_I n_J}} \Biggl\{ 2 \Biggl[
\left( {y'_i \over D^3} \right)_{n0}
\left( {x'_{i'}z'_{j'} \over {D'}^3} \right)_{0n}
- \left( {x'_i \over D^3} \right)_{n0}
\left( {y'_{i'}z'_{j'} \over {D'}^3} \right)_{0n}
\Biggr] \nonumber \\ && \  - 3 \Biggl[ 
\left( {x'_ix'_j ({\bf D} \cdot {\bf a}_I) \over D^5 } \right)_{n0}
\left( {x'_{i'}y'_{j'} \over {D'}^3} \right)_{0n}
- \left( {x'_iy'_j ({\bf D} \cdot {\bf a}_I) \over D^5 } \right)_{n0}
\left( {x'_{i'}x_{j'} \over {D'}^3} \right)_{0n} \nonumber \\ && \ 
+ \left( {y'_ix'_j ({\bf D} \cdot {\bf a}_I) \over D^5 } \right)_{n0}
\left( {y'_{i'}y'_{j'} \over {D'}^3} \right)_{0n}
- \left( {y'_iy'_j ({\bf D} \cdot {\bf a}_I) \over D^5 } \right)_{n0}
\left( {y'_{i'}x'_{j'} \over {D'}^3} \right)_{0n} \Biggr] \Biggr\} + \dots \ ,
\end{eqnarray}
where now $n$ is shorthand for $n_i,n_j$, ${\bf D} \equiv {\bf D}_{ij}$,
${\bf D}' \equiv {\bf D}_{i'j'}$, and the dots represent terms we dropped
which do not contribute within the approximation we invoke in which
the relevant excited states are strictly localized. (However, our
results can be generalized to allow the excited states to extend over a
small complex of atoms, if one simply lets the indices label electrons
in complexes rather than those on atoms.) For localized excited
states, all matrix elements are diagonal in their site indices.
Nonlocal corrections to our results will be small in the parameter $t/E$,
where $t$ is a hopping matrix element which sets the scale of the
width of the band of excited states and $E$ is a typical energy of
the excited states relative to the ground state.\cite{LOCALIZED}
However, it is important to check that these nonlocal corrections
are not proportional to a lower power of $1/R_{IJ}$ than the local
ones we keep.  An analysis of the relative importance of nonlocal
terms is given in Appendix \ref{LOCAL}, where we show explicitly
(albeit only for typical terms when both molecules are excited in
the virtual state) that nonlocal contributions to the chiral
interaction occur at the same order in $1/R_{IJ}$ as do the local
ones, but they are smaller by a factor of order $t/E$.  This
result justifies our subsequent neglect of nonlocal effects.

If both molecules are
in excited states in the virtual state ``$n$," we may set $i=i'$ and
$j=j'$.  If only one molecule, say the $I$th one, is excited in
the virtual state, then $j$ and $j'$ may be different.  We will
consider these two cases in the next two sections.

\section{TWO--MOLECULE TERMS}

In this section we carry the sum in Eq. (\ref{KAPPAEQ}) over
excited states $|i, n; j,m\rangle$, in which molecule $I$
is in state $|i_n\rangle$ with its $i$th atom excited
to its $n$th state and molecule $J$ is in state $|j_m\rangle$
with its $j$th atom excited to its $m$th state.
Because we are dealing with localized states, these virtual
states are obtained from the ground state only by interactions
involving electronic charges on atoms $i$ of molecule $I$
and $j$ of molecule $J$.  Thus we no longer need consider here
the presence of positive nuclear charges.
Neglecting contributions of relative order $(t/ E)$ (as
discussed in Appendix \ref{LOCAL}),
we may set $i=i'$ and $j=j'$ in Eq. (\ref{KAPPAEQ}), so that
the contribution from virtual states in which both molecules
are excited, indicated by the superscript (2), is
\begin{eqnarray}
\label{KAPPA2EQ}
{\tilde \kappa_{IJ}^{(2)} } &=& \sum_{ij} {\sum_n}'
\frac{e^4}{E_{n_In_J}} \Biggl\{ 2 \Biggl[
\left( {y'_i \over D^3} \right)_{0n}
\left( {x'_iz'_j \over D^3} \right)_{n0}
- \left( {x'_i \over D^3} \right)_{n0}
\left( {y'_iz'_j \over D^3} \right)_{0n}
\Biggr] \nonumber \\ && \  - 3 \Biggl[ 
  \left( {x'_ix'_j ({\bf D} \cdot {\bf a}_I) \over D^5 } \right)_{n0}
  \left( {x'_iy'_j \over D^3} \right)_{0n}
- \left( {x'_iy'_j ({\bf D} \cdot {\bf a}_I) \over D^5 } \right)_{n0}
  \left( {x'_ix'_j \over D^3} \right)_{0n} \nonumber \\ && \ 
+ \left( {y'_ix'_j ({\bf D} \cdot {\bf a}_I) \over D^5 } \right)_{n0}
  \left( {y'_iy'_j \over D^3} \right)_{0n}
- \left( {y'_iy'_j ({\bf D} \cdot {\bf a}_I) \over D^5 } \right)_{n0}
  \left( {y'_ix'_j \over D^3} \right)_{0n} \Biggr] \Biggr\} \ .
\end{eqnarray}

We now evaluate this expression using the procedure of Eq. (\ref{MATELEQ}).
To illustrate the calculation for the first
two terms of Eq. (\ref{KAPPA2EQ}) we write
\begin{eqnarray}
T_1 & \equiv & \left( y'_i D^{-3} \right)_{0n} \left( x'_i (\overline z'_j
+ \Delta z'_j) D^{-3} \right)_{n0} - \left( x'_i D^{-3} \right)_{0n}
\left( y'_i (\overline z'_j + \Delta z'_j) D^{-3} \right)_{n0} \nonumber \\
&=& \left( y'_i D^{-3} \right)_{0n} \left( x'_i
(\Delta z'_j) D^{-3} \right)_{n0} - \left( x'_i D^{-3} \right)_{0n}
\left( y'_i (\Delta z'_j) D^{-3} \right)_{n0} \ .
\end{eqnarray}
Expanding the other matrix elements in accord with
Eq. (\ref{MATELEQ}) and recalling that both molecules are
excited in the virtual state, we obtain
\begin{mathletters}
\label{EXPAND}
\begin{eqnarray}
\left( y'_i D^{-3} \right)_{0n} &=&
\left( \Delta {\bf r}'_{i\alpha} \Delta z'_j \right)_{0n}
\nabla'_{i\alpha} \nabla'_{jz} (\overline y'_i \overline D^{-3} )
\nonumber \\
&=& ( \Delta y'_i \Delta z'_j )_{0n} \nabla'_{jz} \overline D^{-3} +
(\Delta z'_i \Delta z'_j )_{0n} \overline y'_i [ \nabla'_{iz}
\nabla'_{jz} \overline D^{-3}] \ .
\end{eqnarray}
Here we have dropped matrix elements like $( \Delta r'_{i \alpha}
\Delta r'_{i\beta})_{0n}$ which involve higher than dipole
excitations and would therefore vanish for the $p$ symmetry we have
assumed for the low--lying excited states.  In any event, since
$\Delta {\bf r} \sim a_0$, the Bohr radius, this term would be smaller
than those we keep by a factor of order $(a_0/R)$.  To write the
second line, one notes that $D$ depends on
${\bf r}'_i$ and ${\bf r}'_j$ only through $z'_i$ and $z'_j$.
Similar relations hold for the other terms in $T_1$, e. g.
\begin{eqnarray}
\left( x'_i \Delta z'_j D^{-3}\right)_{n0} & = &
(\Delta x'_i \Delta z'_j)_{n0} \overline D^{-3}
+ \overline x'_i \left( \Delta z'_i \Delta z'_j \right)_{n0}
\nabla'_{iz} \overline D^{-3} \\
\left( x'_i D^{-3}\right)_{0n} & = &
(\Delta x'_i \Delta z'_j)_{0n} \nabla'_{jz} \overline D^{-3}
+ \overline x'_i \left( \Delta z'_i \Delta z'_j \right)_{0n}
\nabla'_{iz} \nabla'_{jz} \overline D^{-3} \\
\left( y'_i \Delta z'_j D^{-3}\right)_{n0} & = &
(\Delta y'_i \Delta z'_j)_{n0} \overline D^{-3}
+ \overline y'_i \left( \Delta z'_i \Delta z'_j \right)_{n0}
\nabla'_{iz} \overline D^{-3} \ . 
\end{eqnarray}
\end{mathletters}
Thereby for the first two terms of Eq. (\ref{KAPPA2EQ}) we obtain
\begin{eqnarray}
T_1 =  && \Biggl[ \overline x'_i
\left( \Delta y'_i \Delta z'_j \right)_{0n}
\left( \Delta z'_i \Delta z'_j \right)_{n0}
- \overline y'_i \left( \Delta x'_i \Delta z'_j \right)_{0n}
\left( \Delta z'_i \Delta z'_j \right)_{n0} \Biggr]
\nonumber \\ && \times
\Biggl[ \left( \nabla'_{zi} \overline D^{-3} \right)
\left( \nabla'_{zj} \overline D^{-3} \right) - \overline D^{-3}
\left( \nabla'_{zi} \nabla'_{zj} \overline D^{-3} \right) \Biggr] \ .
\end{eqnarray}
Treating the other terms in Eq. (\ref{KAPPA2EQ}) similarly, we obtain
\begin{eqnarray}
\label{KTEQ}
\tilde \kappa_{IJ}^{(2)} &=& 3 \sum_{i , j } (e^4/
\overline D_{ij}^8) [{\bf a}_I \cdot {\bf a}_J
-2({\bf a}_I \cdot \overline {\bf D}_{ij})
({\bf a}_J \cdot \overline {\bf D}_{ij} )/{\overline D}_{ij}^2 ]\nonumber\\
&\times & \Biggl\{ \sum_{\mu ,\nu } E_{\nu \mu}(i,j)^{-1} \Biggl[
\overline y'_j \langle 0_j | \Delta z'_j | \mu_j \rangle
\langle \mu_j | \Delta x'_j | 0 \rangle
- \overline x'_j \langle 0 | \Delta z'_j | \mu_j \rangle \langle \mu_j
| \Delta y'_j | 0 \rangle
\Biggr] \nonumber \\ && \
\times \Biggl[ 2 \langle 0 |\Delta z'_i|\nu_i \rangle^2
- \langle 0 |\Delta x'_i| \nu_i \rangle^2
- \langle 0 |\Delta y'_i| \nu_i \rangle^2 \Biggr] \Biggr\} \ ,
\end{eqnarray}
where $\mu$ and $\nu$ range over the
labels $x$, $y$, and $z$ of the local atomic excited $p$ states and
$E_{\nu \mu}(i,j)$ is the energy of the virtual state relative to the ground
state.  (In principle, this energy can depend on the positions of the
excited atoms.  However, in our simplified treatment we will neglect such
dependence.)  In addition  note that the expression given above for
$\tilde \kappa_{IJ}^{(2)}$ must be averaged with respect to up and
down directions of $I$-th and $J$-th molecules, as in Eq. (\ref{AAVE}).
If the excited states have a degeneracy with respect
to spin, then the sum over $\mu$ and $\nu$ should be extended to
include a sum over spin indices.  However, since singlet--triplet
transitions are nearly forbidden, the multiplicity due to spin does 
not affect our results.  Thus we obtain the result written
in Eq. (\ref{K2RESULT}).

As discussed in the preceding section, our result is similar to that
given by Van der Meer et al.,\cite{Meer} and Kats.\cite{Kats} The important
new aspect of Eq. (\ref{KTEQ}) is that $\kappa^{(2)}_{IJ}$ is expressed
as a sum of contributions from pairs of atoms, one on each
molecule.  This formulation is consistent with the
concept of local chiral centers.\cite{CCENTER}  For $L \ll R$ our expression
for $\kappa^{(2)}_{IJ}$ based on Eq. (\ref{KTEQ}), when written in the
form of Eq. (\ref{K2RESULT}),
reduces to that of van der Meer\cite{Meer} and Kats\cite{Kats} when 
$M_{ij}$ does not depend on $i$ and $j$.
However, when $L$ is not much less than $R$, the fact that
$S_{ij}$ involves an average over distances between
atoms (rather than simply the distance between the centers of
mass of the two molecules), leads to very different results.
In any case, it is important to realize that $M_{ij}$ should
be evaluated with respect to localized states, as is done here.

\subsection{Helical Molecule}

In this subsection we give a concrete evaluation of the above expression
for two identical helical molecules.  In the above formulae, position
operators are given in the coordinate
system fixed in the molecule while matrix elements are taken with
respect to atomic $p$ states which are referred to the principal axes
locally defined for each atom of a molecule. Let us introduce the
parametric representation of coordinates of an atom on a helical
molecule:
\begin{equation}
\label{HELIXEQ}
z'=s,\, x'=a\, \cos(qs),\, y'=a\, \sin(qs),
\end{equation}
where $q$, the chiral wave vector of the helix, is defined so
that a right--handed molecule\cite{HANDED} has $q$ positive.
The locally defined principal axes for the $i$th atom at $z'=s$ 
are chosen in the following way (see Fig. \ref{ATOMF}):
\begin {eqnarray}
\label{LOCALEQ}
{\bf e}''_{ix} &=&\cos(qs) {\bf e}'_x + \sin(qs) {\bf e}'_y \nonumber\\
{\bf e}''_{iy} &=& c\,(- \sin(qs) {\bf e}'_x + \cos(qs) {\bf e}'_y -aq
{\bf e}'_z ) , \nonumber\\
{\bf e}''_{iz} &=& c\,(-aq \sin(qs){\bf e}'_x + aq \cos(qs){\bf e}'_y
+ {\bf e}'_z ) \ ,
\end{eqnarray}
where $c^2=[1+(aq)^2)]^{-1}$.  Here ${\bf e}''_{iz}$ is
the tangent vector to the helix at $z'=s$, ${\bf e}''_{ix}$
is a unit vector along the radius of curvature at $z'=s$, and
${\bf e}''_{iy}$ is the unit vector along the binormal or
the third orthogonal direction.\cite{WIDDER}  We assume that
the principal axes for excited $p$ states coincide with these
principal geometric directions.  If we write
${\bf e}''_{i \mu} = {\cal O}_{i;\mu \nu} {\bf e}'_{i\nu}$, then
the inverse transformation is
${\bf e}'_{i \nu} = {\cal O}_{i;\mu \nu} {\bf e}''_{i\mu}$.

\begin{figure}
\psfig{figure=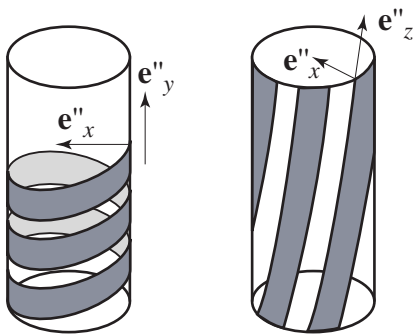}

\vspace{0.2 in} \noindent
\caption{Locally defined principal axes for weakly chiral
molecules with large $q$ (left) and small $q$ (right).
Note that the axis nearly collinear with the long axis of the
molecule is the $y$--axis for large $q$ and the $z$--axis 
for small $q$.  In Eqs. (\ref{SMALLQ}) and (\ref{LARGEQ})
the anisotropy of the polarizability needed is with
respect to the long axis of the molecule.}
\label{WEAK}
\end{figure}

Note that the local axes are defined so that the matrix elements
in Eq. (\ref{KTEQ}) are
\begin{eqnarray}
\langle \mu_i | \Delta {\bf r}'_\nu | 0 \rangle = {\cal O}_{i; \rho \nu} 
\langle \mu_i | \Delta {\bf r}''_\rho | 0 \rangle \ ,
\end{eqnarray}
where $\langle \mu_i | \Delta {\bf r}''_\rho | 0 \rangle$
is nonzero only
if $\rho=\mu$.  Thus, in terms of local atomic coordinates we may
evaluate Eq. (\ref{K2FORMULA}) to obtain
\begin{eqnarray}  
\label{RESULT2}
M_{ij}&=& 3 e^4 c^2 a^2q
\Biggl[ \langle x_i |\Delta x''_i | 0 \rangle^2
\left( \langle z_j |\Delta z''_j |0 \rangle^2 /E_{xz}
- \langle y_j |\Delta y''_j |0 \rangle^2/E_{xy}\right) \nonumber\\
&& + c^2 [ 1 - 2a^2q^2 ] \langle y_i |\Delta y''_i |0\rangle^2
\left( \langle z_j |\Delta z''_j |0 \rangle^2 /E_{yz}
-\langle y_j |\Delta y''_j |0 \rangle^2/ E_{yy} \right) \nonumber\\
&& + c^2 [  2 - a^2q^2 ] \langle z_i |\Delta z''_i |0 \rangle^2
\left( \langle y_j |\Delta y''_j |0 \rangle^2 /E_{zy}-\langle z_j
|\Delta z''_j |0\rangle^2/E_{zz} \right) \Biggr] \ .
\end{eqnarray}
This quantity can not depend on the locations of sites $i$ and $j$
because it is invariant against rotation about the long axis of the
molecule and all locations on the helix are equivalent once end effects
are neglected.  Thus, neglecting end effects, we obtain the limiting results,
\begin{mathletters}
\begin{eqnarray}
\label{SMALLQ}
M_{ij} & \equiv & M \approx 6e^4 a^2 q \Biggl\{
\langle y|\Delta y'' |0 \rangle^2
\Biggl( { \langle z|\Delta z'' |0 \rangle^2 \over E_{z y}}
-\frac{\langle x|\Delta x'' |0 \rangle^2} {2E_{x y}}
-\frac{\langle y|\Delta y'' |0 \rangle^2} {2E_{y y}} \Biggr)
\nonumber\\ &-&\langle z|\Delta z'' |0 \rangle^2 \Biggl(
\frac{\langle z|\Delta z'' |0 \rangle^2} {E_{z z}}
-\frac{\langle x|\Delta x'' |0\rangle^2} {2E_{x z}}
-\frac{\langle y|\Delta y'' |0\rangle^2} {2E_{y z}} \Biggr) \Biggr\} 
\ ,  \ \ \ (aq)^2 \ll 1 \ ; \\
\label{LARGEQ}
& \approx & 6 {e^4 \over q} \Biggl\{
\langle y |\Delta y'' |0\rangle^2
\Biggl( \frac{\langle y |\Delta y'' |0 \rangle^2} {E_{y y}}
-\frac{\langle x|\Delta x'' |0 \rangle^2} {2E_{x y}}
-\frac{\langle z|\Delta z'' |0 \rangle^2} {2E_{z y}} \Biggr)
\nonumber\\ &-&\langle z|\Delta z'' |0 \rangle^2 \Biggl(
\frac{\langle y|\Delta y'' |0 \rangle^2} {E_{y z}}
-\frac{\langle x|\Delta x''|0 \rangle^2} {2E_{x z}}
-\frac{\langle z|\Delta z''|0 \rangle^2} {2E_{z z}}
\Biggr) \Biggr\} \  , \ \ \ (aq)^2 \gg 1 \ . 
\end{eqnarray}
\end{mathletters}
In both limits, the molecule is only weakly chiral, as we illustrate
in Fig. \ref{WEAK}.  (To measure
chiral strength the criterion of Ref. \onlinecite{HKL} may be invoked.)

Now let us consider $\kappa^{(2)}_{IJ}$ as a function of the molecular
length $L$.  For simplicity we assume that the molecules are aligned
exactly along their local nematic directions.  Also we simplify
the calculation by considering only the case when ${\bf R}_{IJ}$
is perpendicular to ${\bf a}_I$.  Thus we will set
\begin{eqnarray}
\label{EQUATOR}
{\bf a}_I \cdot {\bf a}_J =1,\ \ \ \ \ \
(\overline {\bf D}_{ij} \cdot {\bf a}_I)
(\overline {\bf D}_{ij} \cdot {\bf a}_J) =
(\overline z_j - \overline z_i )^2 \ , \ \ \ \ \ 
\overline D_{ij}^2 = R^2 + (\overline z_i - \overline z_j )^2 \ .
\end{eqnarray}
Then
\begin{eqnarray}
\label{S2EQ}
\kappa^{(2)}_{IJ} & = & M
\sum_{ij} (1/\overline D_{ij}^8) [{\bf a}_I \cdot {\bf a}_J
-2 (\overline {\bf D}_{ij} \cdot {\bf a}_I)
(\overline {\bf D}_{ij} \cdot {\bf a}_J ) / \overline D_{ij}^2 ]
\nonumber \\
& = &  {N^2 M \over L^2 } \int _{-L/2}^{L/2} \int_{-L/2}^{L/2} dz_I dz_J \, 
\frac{R^2-(z_I-z_J)^2}{[R^2+(z_I-z_J)^2]^5} \nonumber \\ &=&
{ \rho^2 M \over R^6}
\Biggl[ \frac{15L}{32R} \tan^{-1} \left( \frac{L}{R} \right)
+\frac{L^2(51R^4+72 R^2 L^2+ 29 L^4)}{96(R^2+L^2)^3}\Biggr] \ ,
\end{eqnarray}
where $N$ is the number of atoms in a molecule and $\rho=N/L$
is the number of atoms per unit length in the molecule.  For
this simple calculation the average of Eq. (\ref{AAVE}) is
superfluous, so that $\kappa^{(2)}_{IJ} = \tilde \kappa^{(2)}_{IJ}$.
The asymptotic result for $L\ll R$ that
$\kappa^{(2)}_{IJ} \sim R^{-8}$ can be seen in previous
calculations.\cite{Meer,Kats}  However, even in this limit,
the fact that $\kappa^{(2)}_{IJ}$ is proportional to $L^2$
is not apparent from the previous results.  To our knowledge,
our result that $\kappa^{(2)}_{IJ} \sim L/R^7$ for
$L \geq R$ is a new one.

The macroscopic chiral wave vector ${\bf Q}$ and ${\bf R}$ are both
taken perpendicular to the nematic direction.  For $QR_{IJ} \ll 1$,
we have ${\bf a}_I \times {\bf a}_J \cdot {\bf R}_{IJ} = -QR^2
\cos^2 \phi_R$, where $\phi_R$ is the angle between ${\bf R}_{IJ}$
and ${\bf Q}$.  Then the chiral energy per molecule from virtual
states with two molecules excited, ${\cal E}^{(2)}$, is given by
\begin{eqnarray}
\label{CALE}
{\cal E }^{(2)} & \equiv & \case 1/2 \sum_J \langle E_{IJ} \rangle
= - \case 1/2 \sum_J QR^2 \cos^2 \phi_R \kappa^{(2)}_{IJ}
\nonumber \\ & = & - \case 1/4 \gamma  M \rho^2 (QR)\frac{1}{R^5}
\Biggl[ \frac{15L}{32R} \tan^{-1} \left( \frac{L}{R} \right)
+\frac{L^2(51R^4+ 72R^2 L^2+ 29 L^4)}{96(R^2+L^2)^3}\Biggr] \ .
\end{eqnarray}
In obtaining this result we approximated the sum over $J$ by a sum
over $\gamma$ nearest neighbors in the plane as specified in
Eq. (\ref{EQUATOR}), so that $\cos^2 \phi_R \rightarrow \case 1/2$.
>From the discussion in Appendix \ref{PERP} we are led to believe
that the result of Eq. (\ref{CALE}) will not be seriously
modified by taking a more realistic distribution of nearest
neighboring molecules.  We identify this result with the
contribution to the torque field $h$ in the Frank free energy
from virtual states with two molecules excited:
\begin{mathletters}
\begin{eqnarray}
h^{(2)} & = & - {{\cal E}^{(2)} \over \Omega Q} = {\gamma M \rho^2 L \over
4 \Omega R^5} \Biggl[ \frac{15}{32} \tan^{-1} \left( \frac{L}{R} \right)
+\frac{LR(51R^4+ 72 R^2 L^2+ 29 L^4)}{96(R^2+L^2)^3}\Biggr] \\
& = & {\gamma M \rho^2 L^2  \over 4 R^9 } \ , \ \ \ L \ll R \\
& \approx & {15 \pi \gamma M \rho^2 \over 256 R^7 }
\ , \ \ \ L \geq R \ ,
\end{eqnarray}
\end{mathletters}
where we took the volume per molecule, $\Omega$, to be $\Omega= LR^2$
for $L \geq R$ and $R^3$ for $L \ll R$.  For $L \ll R$,
$h^{(2)} \propto (L^2/R^9)$, consistent with the previous results of Refs.
\onlinecite{Meer} and \onlinecite{Kats},
whereas for $L \geq R$, $h^{(2)} \propto 1/R^7$.
 
\subsection{Numerical Estimate of the Macroscopic Pitch}

Now we want to estimate the value of the pitch using the result
for ${\cal E}$ obtained above.  Intuitively one expects that the 
polarizability tensor will have its largest component tangent
to the helix and that the anisotropy of the polarizability in the
plane perpendicular to the tangent will be small.  Essentially,
we will attribute the anisotropy of the polarizability to
the anisotropy in the excitation energy $E_\alpha$.  Therefore,
somewhat arbitrarily, we will take all the matrix elements like
$|\langle \mu|\Delta {r''}_\alpha |0\rangle |$, where $\mu=x,y,z$, to
have same value, $a_a$,  where $a_a$ is of order the radius of an atom.
We therefore parametrize the excitation energies
in Eq. (\ref{RESULT2}) as
\begin{eqnarray}
\label{DELTAEQ}
E_x/E = 1 + \case 1/3 \delta + \eta \ , \ \ \ \
E_y/E = 1 + \case 1/3 \delta - \eta \ , \ \ \ \
E_z/E = 1 - \case 2/3 \delta \ ,
\end{eqnarray}
where $E$ is the average excitation energy.  Within our assumption of
constant matrix elements the parameters $\delta$ and $\eta$ characterize
the anisotropy of the excitation energy and through it the anisotropy
of the atomic polarizability.  When this anisotropy is small, we find that
\begin{eqnarray}
\label{KIJEQ}
M &=& - {3 e^4 a_a^4 a \over 2E} \left( {a q \over 1 + a^2 q^2 } \right)
(\delta- \eta) \Psi(aq) \equiv - {3 e^4 a_a^4 a \over 2E} 
G(\delta, \eta, aq ) \ ,
\end{eqnarray}
where
\begin{equation}
\label{PSIEQ}
\Psi(aq) = { \delta - \case 1/2 (aq)^2 (\delta - 3 \eta)
\over (1+ a^2q^2) }
\end{equation}
and
\begin{eqnarray}
G(\delta, \eta, aq) = \left( { aq (\delta - \eta) \over
1+ a^2 q^2 } \right) \Psi (aq)  \ .
\end{eqnarray}
The corresponding results for $\kappa^{(2)}_{IJ}$ are
\begin{mathletters}
\begin{eqnarray}
\kappa^{(2)}_{IJ} &=& - {3 e^4 a_a^4 a \rho^2 L^2 \over
2 E R^8 } G(\delta, \eta, aq) \ , \ \ L \ll R \\
& \approx  &  - {45 \pi e^4 a_a^4 a \rho^2 L \over 128 E R^7}
G(\delta , \eta, aq) \ , \ \ L \geq R \ .
\end{eqnarray}
\end{mathletters}

\begin{figure}
\psfig{figure=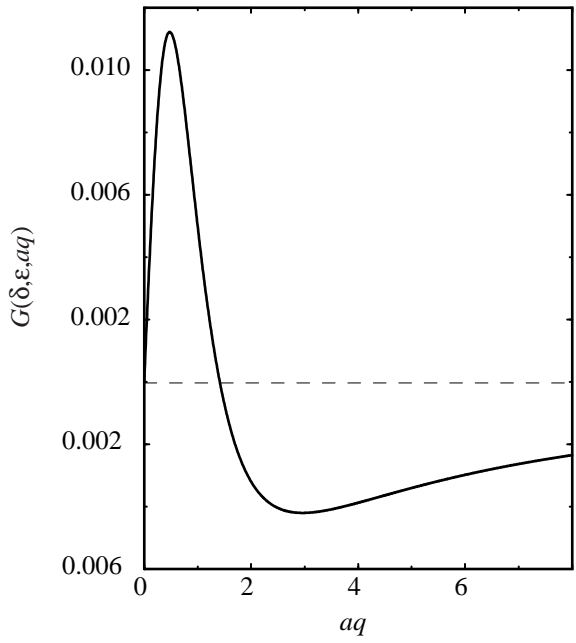}
\vspace{0.2 in} \noindent
\caption{
The function $G(\delta, \eta, aq)=\left( {aq \over 1 + a^2 q^2 } \right)
(\delta - \eta) \Psi(aq)$ versus $aq$ for $\delta = \case 1/5$ and $\eta =0$.}
\label{GFIG}
\end{figure}

Thus $M$ is quadratic in the anisotropy of the polarizability and
\begin{mathletters}
\label{HRESULT}
\begin{eqnarray}
h^{(2)} & = &  - {3 \gamma e^4 a_a^4 a \rho^2 L^2 \over 8ER^9}
G(\delta , \eta ,aq) \ , \ \ L \ll R \\
& \approx & - {45 \pi \gamma e^4 a_a^4 a \rho^2 \over 512 ER^7 }
G(\delta , \eta ,aq ) \ , \ \ L \geq R \ .
\end{eqnarray}
\end{mathletters}
This conclusion is a natural one: surely the torque field must disappear
when the anisotropy of the polarizability is turned off.  Also,
when $E_z = E_y$ (i. e. when $\delta=\eta$), the chiral constant
${\sigma'}$ vanishes.  To see that note that when $E_z=E_y$,
one of the principal axes for each atom can be taken to
be perpendicular to the axis of the helix, in which case
the matrix elements appearing in $\sigma'$ are
invariant with respect to the mirror operation $z' \rightarrow - z'$.
To illustrate the dependence of $h^{(2)}$ on the molecular
chiral wavevector $q$, we show in Fig. \ref{GFIG} $G(\delta, \eta, aq)$
versus $aq$ for fixed values of $\delta $ and $\eta$.  There
one sees that $h^{(2)}$ is maximal for $aq$ of order unity
and decreases rapidly away from this maximum.  Of course, an
experimental test of this dependence is difficult since varying
$q$ at constant $\rho$ involves structural changes in a molecule.
To treat small chirality we take $aq=1/3$ (or $aq=3$) and
we set $a_a=1{\AA}$, $E=8$ eV (these parameters correspond to an atomic 
polarizability $\alpha =2e^2a_a^2/E=27a_0^3$),
$a=7.5{\AA}$, $\gamma =6$,  $L=200 {\AA}$,
$R=20 {\AA}$, $\rho =3 {\AA}^{-1}$, $\delta= 1/5$, and $\eta=0$.
With the volume per molecule, $\sim LR^2$, the chosen values of
the parameters correspond to volumetric density of molecules
of about $40\% $ and a dielectric constant,
$\epsilon  = 1 + 4 \pi \alpha \rho L / \Omega \approx 1.3$.
Then the torque field is approximately
$h = 4.5\times 10^{-4}$ (dyne/cm).  If now one takes the
Frank constant $K_2$ to be $10^{-7}$dyne,
then the macroscopic pitch of the liquid crystal will be
$P=2\pi /Q= 2 \pi K_2/h =-14 \mu$  (or $28\mu$ for $aq=3$).
If we had taken $\delta=3/10$ and $\eta=0$, then the pitch
would be $-4.5 \mu$ (or $9.0 \mu$ for $aq=3$).

It may be seen that the computed pitch is longer then one
usually finds experimentally for a system consisting of molecules
of the above size.  There are two possible explanations for
this discrepancy.  First of all, our approximations, although
improved over previous ones, may still not be sufficiently
accurate.  For example, for two helices of radius 7.5 $\AA$
at a center--to--center separation of 20 $\AA$ have their
nearest groups separated by only $5\AA$.  Under these
conditions, the expansion in terms of even the transverse
coordinates of the atoms may not be rapidly convergent.
The second possible reason for the discrepancy between
calculated and observed pitches would be that an
explanation of the pitch of cholesterics requires
consideration of steric interactions.  We are presently
considering how our arguments might be improved to
discriminate between these two explanations.

If one can find molecules for which quantum chiral
interactions considered in this section are dominant, then the
following remarks are relevant.
Notice that for helical molecules the torque field, $h$, can
have either sign in both the large $q$ and small $q$ limit,
depending on the signs of $(\delta - \eta)$ and $(\delta - 3 \eta)$.
This is in contrast to the situation for steric interactions,
for which it is believed\cite{STERIC} that the contribution to $h$
from the repulsive (i. e. steric) chiral interaction between molecules
is negative for small $q$ and is positive for large $q$.
Helical molecules which do not follow the sign prediction for $h$ due
to repulsive steric interactions might constitute examples of
molecules for which the quantum dispersion forces dominate the
chiral interactions.  In general, the density dependence of the
quantum and steric contributions to $h$ will be different.  Thus,
if these two mechanisms compete, it is likely that the sign of
$h$ could depend on the density.

\section{ONE--MOLECULE TERMS} 

In the model of a molecule considered before we supposed it to
consist of He--like atoms. In reality one would expect the
outer electronic shell of atoms to be deformed by the interaction
with nearest neighbors. In general, constituent atoms or complexes
will possess a dipole moment. Hence it is of interest to consider the
situation when one of the molecules is in its ground state
in the virtual state of two--molecule system.  Up to now
this case was ignored, although, as we shall see, it may play
a significant, if not dominant role.

>From Eq. (\ref{KAPPAEQ}) we obtain the following expression for
the additional contribution, denoted $\kappa_{IJ}^{(1)}$, to
$\kappa_{IJ}$ from virtual states in which only one molecule
is excited.  We still invoke the approximation of localized
excited states.\cite{LOCALIZED}  But then terms in which only molecule $J$
is excited require evaluation of ${\cal E}_{ij;i'j'n}$ with
$j=j'$, but $i$ and $i'$ are arbitrary and similarly when only
molecule $I$ is excited.  For a molecule in the
excited state we use the same approximation as before, again
expanding the denominator with respect to $\Delta {\bf r}$ to get
a nonzero matrix element.  For the molecule which remains
in its ground state in the virtual state, one has to include
both signs of charge at each site.  Thus (see Appendix
\ref{h1app}) we find that
\begin{eqnarray}
\label{KAPPA1EQ}
\tilde \kappa_{IJ}^{(1)} & = & 6 \sum_{i,i' \in I; j \in J}
e^2 q_i q_{i'}(\overline x'_i \overline y'_{i'}
- \overline x'_{i'} \overline y'_i)
\frac{ \left( \overline {\bf D}_{i'j} \cdot {\bf a}_J \right) }
{{\overline D}_{ij}^3 {\overline D}_{i'j}^5} \nonumber\\
&\times &\sum_{\mu } E_\mu(j)^{-1}
(\langle \mu_j | \Delta z'_j |0\rangle^2
-\case 1/2 \langle \mu_j | \Delta x'_j |0\rangle^2
-\case 1/2 \langle \mu_j | \Delta y'_j |0 \rangle^2 ) \ .
\end{eqnarray}

In Eq. (\ref{KAPPA1EQ}) we sum $i$ and $i'$ over all the charges in
a given atom, in which case $q_i \overline x'_i$ is replaced by $p'_{xi}$,
where ${\bf p}_i$ is now the expectation value of the dipole moment
of the $i$th atom, in its ground state, so that $i$ and $i'$ from now
on refer to atoms, whereas $j$ will still label electronic charges.
Then the preceding equation can be reduced to
\begin{mathletters}
\label{KAPPA1}
\begin{eqnarray}
\tilde \kappa_{IJ}^{(1)} & = & 6 \sum_{ii' j}
[p'_{ix} p'_{i'y} - p'_{i'x} p'_{iy} ]
\frac{(\overline {\bf D}_{i'j} \cdot {\bf a}_J)}
{\overline D_{ij}^3 \overline D_{i'j}^5} \nonumber\\
&\times e^2 &\sum_{\mu } E_\mu(j)^{-1}
[\langle \mu_j | \Delta z'_j |0 \rangle^2
- \case 1/2 \langle \mu_j | \Delta x'_j |0 \rangle ^2
- \case 1/2 \langle \mu_j | \Delta y'_j |0 \rangle ^2 ] 
\label{KAPPA1A} \\ & \equiv & W_1 W_2 \ ,
\end{eqnarray}
\end{mathletters}
where $W_1$ is the factor on the first line of this
equation and $W_2$ that on the second line.
In writing this result we assumed that for typical atoms, $i$,
one has $p_{ix} \overline z_{i'} \gg p_{iz} \overline x_{i'}$.
Once again, in this expression one has to carry out
averaging with respect to independent up and down
orientations of both molecules.  But this average
turns out to be superfluous for the model of a helical
molecule which was introduced above.

\begin{figure}
\psfig{figure=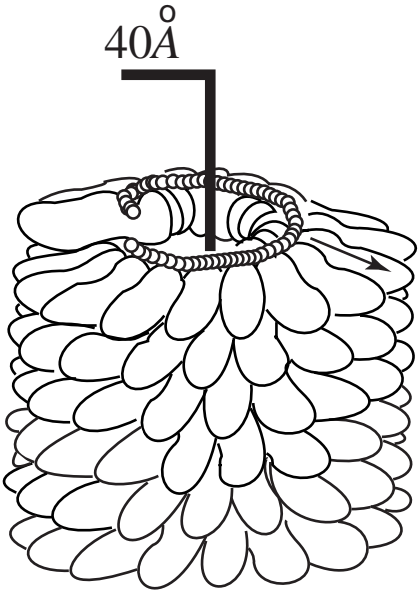}

\vspace{0.2 in} \noindent
\caption{TMV, taken from Ref. \protect\onlinecite{TMVREF}.  We indicate
a possible axis along which the dipole moment of each complex
might be oriented.  In the situation shown here, the largest
component of the dipole moment of the complex is radial.}
\label{TMVFIG}
\end{figure}

As in Eq. (\ref{LOCALEQ}), we introduce components of the
atomic dipole moment with respect to the principal axes of the atom,
in which case we have
\begin{eqnarray}
p'_x&=&p''_x \cos(qs)-cp''_y \sin(qs)-caq p''_z \sin(qs) \nonumber\\
p'_y&=& p''_x \sin(qs)+cp''_y \cos(qs)+caq p''_z \cos(qs) \nonumber\\
p'_z&=&-caq p''_y + cp''_z \ . 
\end{eqnarray} 
The component $p_x''$ is essentially the radial component of the atomic
dipole moment and is nonzero for helical geometry.  For instance, for the
molecule TMV, shown in Fig. \ref{TMVFIG},\cite{TMVREF} this
radial component may be appreciable.  In such a case we write
\begin{mathletters}
\begin{eqnarray}
\wp^2_{ii'} & \equiv  & p'_{ix} p'_{i'y} - p'_{i'x} p'_{iy} = 
[{p''_x}^2+c^2(p''_y + aq p''_z )^2] \sin[q(s_{i'}-s_i)] \\
& \equiv & \wp_0^2 \sin [q(s_{i'}-s_i)] \ .
\end{eqnarray}
\end{mathletters}
We now substitute this form into Eq. (\ref{KAPPA1A}) and assume
perfect alignment as in Eq. (\ref{EQUATOR}).  Then the summand is
symmetrized and we write $W_1= \wp_0^2 X_1 L/R^8$, with
\begin{eqnarray}
\label{X1EQ}
X_1 (\tilde q , \tilde L) & = & 3 \sum_{i i' j}
\left[ { 1 - \tilde L^2 \left( \tilde {s_j}^2 + \tilde s_i \tilde s_{i'}
- \tilde s_j ( \tilde s_i + \tilde s_{i'} ) \right)
\over [1+ \tilde L^2 (\tilde s_j- \tilde s_i)^2]^{5/2}
[1+ \tilde L^2 (\tilde s_j- \tilde s_{i'})^2]^{5/2} }
\right] \left( \tilde s_{i'} - \tilde s_i \right)
\sin[\tilde q (\tilde s_{i'}-\tilde s_i)] \ ,
\end{eqnarray}
where $\tilde s = s/L$, $\tilde q = qL$, and $\tilde L = L/R$.
To evaluate $W_2$ we again invoke the model of Eq. (\ref{DELTAEQ}),
in which case, for small anisotropy, Eqs. (\ref{LOCALTR}) and
(\ref{HELIXEQ}) enable us to write
\begin{eqnarray}
W_2  & \equiv & e^2 \sum_{\mu } E_\mu^{-1}
[\langle \mu_j | \Delta z'_j | 0 \rangle^2
- \case 1/2 \langle \mu_j | \Delta x'_j |  \rangle^2
- \case 1/2 \langle \mu_j | \Delta y'_j |  \rangle^2 ] \nonumber \\
& = & e^2 \left[ -\frac{\langle x | \Delta x'' | 0 \rangle^2} {2E_x}
+ \frac{\langle y | \Delta y'' | 0 \rangle^2} {2E_y} 
\frac{2(aq)^2-1}{1+(aq)^2}
+ \frac{\langle z | \Delta z'' | 0 \rangle^2}
{2E_z} \  \frac{2-(aq)^2}{1+(aq)^2} \right]  \nonumber \\
&=& { e^2 a_a^2 \over E} \Psi(aq) \ ,
\end{eqnarray}
where $\Psi(aq)$ is defined in Eq. (\ref{PSIEQ}).
Using the asymptotic evaluations in Appendix \ref{X}, we thus
have the results
\begin{mathletters}
\label{KAPRESEQ}
\begin{eqnarray}
\tilde \kappa^{(1)}_{IJ} & = & {e^4 a_a^2 d^2 L^4 \rho^3 \over
E R^8 } \Psi(aq) \phi(\case 1/2 qL) \ , \ \ \ a \ll L \ll R \ ; \\
& = & { 8 e^4a_a^2 d^2 \rho^3 q L \over ER^4} \Psi(aq) I_1^2(qR) \ ,
\ \ \  L \gg R \ ;
\end{eqnarray}
\end{mathletters}
where $d$ is the effective size of the dipole moment: $\wp_0=ed$,
\begin{eqnarray}
\label{PHIEQ}
\phi(x) = - (3/2) (d/dx)[(\sin x)/x]^2 \ ,
\end{eqnarray}
 and
\begin{eqnarray}
\label{INTEQ}
I_n(qR) = \int_0^\infty e^{-\case 1/2 [x^2 + (qR/x)^2]} x^n dx \ .
\end{eqnarray}

\begin{figure}
\psfig{figure=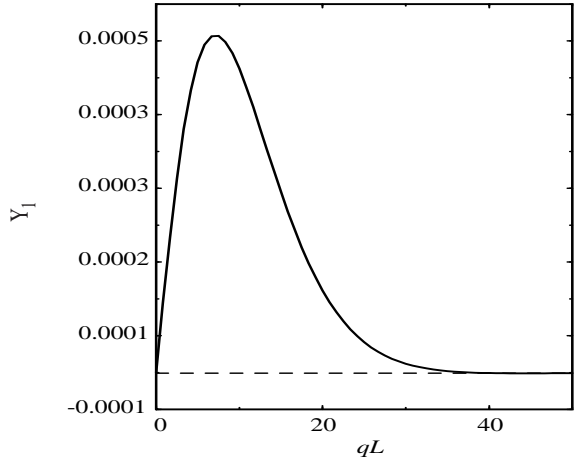,height=3.5in,width=3in}

\vspace {-1 in} \noindent
\caption{$Y_1(qL,L/R)= \Psi (aq) X_1( qL , L/R )$, with  
$ \Psi (aq)$ and   $X_1( \tilde q , \tilde L )$ defined in Eqs. 
(\protect\ref{PSIEQ}) and (\protect\ref{X1EQ}) respectively,
versus $qL$ for $L=200\AA $, $ R=20\AA$, $\rho=3\AA^{-1}$,
$\delta=1/5$, and $\eta =0$ . According to Eq. (\protect\ref{H1EQ})
the quantity plotted gives the dependence of the torque field $h^{(1)}$
on the chiral wavevector of a molecule $q$. Note that the molecule
is achiral if either $q \rightarrow 0$ or $q \rightarrow \infty$.}
\label{ASYMPT}
\end{figure}

\begin{figure}
\psfig{figure=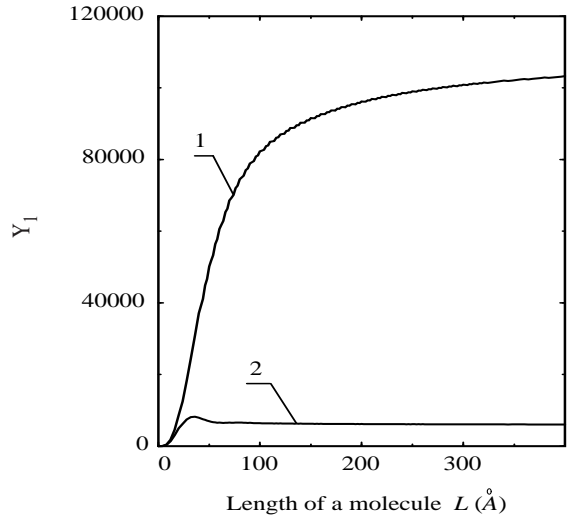,width=4in,height=3in}
\caption{The factor $Y_1\equiv \Psi(aq) X_1 ( qL , L/R)$ 
as a function of $L$ for two values of the  molecule wave number 
$q=0.0444 \AA^{-1}$ (plot 1) and $q=0.1333 \AA^{-1}$ (plot 2)
with $R=20\AA$, $a=7.5\AA$, $\delta=0.2$, and $\eta=0$.
According to Eq. (\protect\ref{H1EQ}) when $L \gg R$ (so that
$\Omega = LR^2$) the quantity plotted gives the dependence of the
torque field $h^{(1)}$ on the length $L$ of a molecule.}
\label{X1FIG}
\end{figure}

Now we evaluate $h$ following the procedure of Eq. (\ref{CALE}) in
terms of the chiral energy per molecule ${\cal E}^{(1)}$ due to
one--molecule effects:
\begin{eqnarray}
\label{H1EQ}
h^{(1)} = - {{\cal E}^{(1)} \over \Omega Q }  =  {\gamma R^2 \over 4\Omega} 
W_1 W_2 = \left( {\gamma e^4 a_a^2 d^2 L \over 4 E R^6 \Omega } \right)
\Psi (aq) X_1( \tilde q , \tilde L ) \ .
\end{eqnarray}
Using the evaluations of Appendix \ref{X}, we obtain the asymptotic results,
\begin{mathletters}
\begin{eqnarray}
h^{(1)} & = & {\gamma e^4 a_a^2 d^2 L^4 \rho^3 \over 4ER^9 }
\Psi(aq) \phi(\case 1/2 qL) \ , \ \ \ \ \ \ a \ll L \ll R \ , \\
& = & {2 \gamma e^4 a_a^2 d^2 \rho^3 q \over ER^4 }
\Psi(aq) I_1^2(qR) \ , \ \ \ \ \ \ L \gg R \ ,
\end{eqnarray}
\end{mathletters}
Here again we see from the appearance of $\Psi(aq)$ that chirality
requires a nonzero anisotropy of the polarizability characterized
by $\delta$ and $\eta$.  Since the factor $\Psi(aq)$ also appears
in Eq. (\ref{HRESULT}), we see that the critical value (if any)
where $h$ changes sign as $q$ is varied, is only
determined by the geometry, at least within our simple model.
For concentrated systems, the limit $L \gg R$ is the most
relevant and for this case Fig. \ref{ASYMPT} shows how
$h^{(1)}$ depends on the molecular chirality $q$, when the length
of the molecules and the density of atoms $\rho$ are fixed. 
Note that the variation of $h^{(1)}$ with the molecular chirality $q$
strongly depends on details of molecular geometry since only a fixed number 
of atoms is allowed on a helical thread. Figure \ref{ASYMPT} shows
that for $aq$ of order or less than unity, where $h^{(1)}$ is
appreciable and may give a short pitch, $h^{(1)}$
is positive, whereas for steric interactions $h$ is believed to
be negative for small $q$.\cite{Straley,STERIC}
Since increasing the density probably
causes steric interactions to dominate, it is possible
that the combination of these two mechanisms could cause $h$
to change sign as a function of density or temperature.\cite{PREIS}
In Fig. \ref{X1FIG}  we show the behavior of the quantity
$Y_1 \equiv \Psi(aq) X_1( \tilde q , \tilde L )$ as a function
of $L$ for $ R=20\AA$, $\rho=3\AA^{-1}$, $\delta=1/5$, and $\eta =0$
for two fixed values of the molecular chiral wavevector, $q$.
In particular, it is noteworthy that for large $L$, $Y_1$
(and therefore $h^{(1)}$) is independent of $L$.
To get some idea of the relative importance of $h^{(1)}$ and
$h^{(2)}$, consider their ratio:
\begin{mathletters}
\begin{eqnarray}
r \equiv {h^{(1)} \over h^{(2)}} 
& = & - \left( {2L^2d^2 \rho \over 3 a_a^2 a } \right)
\left( {1 + a^2 q^2 \over aq} \right) (\delta - \eta)^{-1}
\phi(\case 1/2 qL) \ , \ \ \ L \ll R \ ; \\
& = & - \left( {1024 R^3 d^2 \rho \over 45 \pi a_a^2 a^2 } \right)
( 1 + a^2 q^2 ) (\delta - \eta)^{-1} I_1^2(qR)  \ ,\ \ \ L \gg R \ .
\end{eqnarray}
\end{mathletters}
One sees that even with $d/a_a$ as small as 0.03, this ratio can easily
be of order unity.

To numerically estimate the pitch arising from the considered
interaction we will take parameters of a system and constituent 
molecules chosen in preceding section. Then, if $aq=1/3$ one finds 
$h^{(1)}=0.5(d/a_0)^2$ dyne/cm. If molecules 
posses a local dipole moment, the resulting dipolar interactions
may lead to strong biaxial correlations between neighboring
molecules.  Using the evaluation of the dipolar interaction energy
in terms of the integral analyzed in Appendix \ref{X}, we estimate
that the order of magnitude of the dipole--dipole interaction to
be $V_{dd} \approx (d/a_0)^2\cdot 10^5\, $K.
So if we suppose that the biaxial correlations due to dipole-dipole
interaction among molecules is negligible when it is less than $100 K$ then
one must have $(d/a_0)^2 < 10^{-3}$. At the upper limit of validity of
our calculations $(d/a_0)^2 = 10^{-3}$ and the
macroscopic pitch due to $h^{(1)}$ will be
$P^{(1)}=2\pi K_2/h^{(1)}=12.5\, \mu$.   As the density of local 
dipoles is increased, the macroscopic pitch becomes smaller.
For instance, if we set $d/a_0=1/3$, we get a pitch of order
0.1$\mu$, although this estimate will be significantly modified
by biaxial correlations, which have been neglected in our treatment.
Since $h^{(1)}\sim \rho^3$ and the dipole--dipole interaction is
proportional to $\rho^2$, it is conceivable that for much larger
molecules $h^{(1)}$ could be significant without
the dipoles being large enough to induce long range biaxial order.
Finally, when $aq$ is larger than unity (as for TMV), this mechanism
leads to a very large pitch for almost any choice of parameters.
As mentioned in Sec. IVB, it is possible that larger values
of the pitch would be obtained if the role of the transverse
were treated exactly rather than by an expansion.

%%%%%%%%%%%%%%%%%%%%%%%%%%%%%%%%%%%%%%%%%%%%%%%%%%%%%%%%%%%%%%%%%%%%
%     CONCLUSION
%%%%%%%%%%%%%%%%%%%%%%%%%%%%%%%%%%%%%%%%%%%%%%%%%%%%%%%%%%%%%%%%%%%%
	
\section{Conclusion}

Here we put our work into the context of current research and
record our conclusions.

\vspace{0.2 in} \noindent
1)  We introduced a simple model of localized polar excited states
that enabled us to make an explicit calculation of the chiral
interaction, $\kappa_{IJ} {\bf a}_I \times {\bf a}_J \cdot {\bf R}_{IJ}$,
between molecules $I$ and $J$ due to quantum charge fluctuations
analogous to those responsible for the $R^{-6}$ dispersion 
interaction between neutral atoms.  We identified two 
distinct physical effects depending on whether one or both
molecules were excited in the virtual state of the two--molecule
system.  In implementing this calculation we used a modified
multipole expansion in which only coordinates transverse to the
long axis of the molecule were expansion parameters, so that
we could treat long molecules which usually are the building blocks 
of liquid crystals.  The contribution, $\kappa^{(2)}_{IJ}$, to
$\kappa_{IJ}$ from virtual states with both molecules excited has a
form similar to that found by van der Meer et al\cite{Meer} and
Kats\cite{Kats}.  For a helical molecule of length $L$ we find
that $\kappa^{(2)}_{IJ} \propto L^2/R^8$ for $L \ll R$ and
$\kappa^{(2)}_{IJ} \propto L/R^7$, for $L \ge R_{IJ}$.  The
contribution, $\kappa^{(1)}_{IJ}$ to $\kappa_{IJ}$ from virtual
states with only one molecule excited is usually only dominant
when the local atomic dipole moments are large enough to give
rise to significant (possibly long--range) biaxial correlations.
Both mechanisms give rise to a chiral interaction between a
chiral molecule and an achiral one that has a local
anisotropic polarizability.  Our formulation leads to numerical
estimates of the pitch which are larger than that found
in many cholesterics.  Whether this discrepancy is an
artifact of the expansion in transverse coordinates 
along with a disregard of biaxial correlations between molecules
or is  an indication that steric rather than quantum interactions
are the microscopic origin of macroscopic chirality is
not clear at present.  The role of biaxial correlations
between molecules will be considered elsewhere.\cite{IH}

\vspace{0.2 in} \noindent
2)  We evaluated $\kappa_{IJ}$ and the torque field $h$ for helical
molecules as a function of the wave vector $q$ which describes the
chiral structure of an individual molecule.  We found that the sign
of $h$ depends on the details of the anisotropy of the local atomic
polarizability.  For instance, for $(aq)^2 \ll 1$, the sign of
$h^{(1)}$ (the contribution to $h$ from virtual states in which
only one molecule is excited) has the same sign as $\delta$,
the local anisotropy of the polarizability. One expects $\delta$
to be positive because presumably the polarizability along the
tangent of the helix is larger than that along perpendicular
directions.  This sign of $h^{(1)}$ is opposite to that expected
from steric interactions.\cite{STERIC} As for steric interactions,
one expects $h^{(1)}$ to change sign as $q$ is increased, but
our calculations indicate that this only happens when $h^{(1)}$ is
so small that it is hardly likely to be the dominant mechanism for
macroscopic chirality.   When $\delta$ is positive and large,
the sign of the two--molecule contributions to $h$ is negative
for small $aq$ and positive for large $aq$, just as expected for
steric interactions.  However, our calculations indicate that
normally $h^{(2)}$ is not significant.

\vspace{0.2 in} \noindent
3)  Here we calculated the effective chiral interactions
by averaging the orientation of the molecule over configurations
with the long axis fixed.  Even within mean field theory,
wherein each molecule is described by a single--molecule
orientational distribution function of the three Euler angles,
the only required symmetry in the locally nematic state is
that it be invariant against rotations about the nematic axis.
As discussed in Appendix B, this requirement still permits
biaxial contributions to the orientational probability
distribution which we neglected.

\vspace{0.2 in} \noindent
4)  These calculations suggest some general observations.  First
of all, the interaction from virtual states with two molecules
excited, give rise to a two--point chiral interaction in the form
of an integral over the long axis of each molecule.  This result
gives a formal justification for introduction of a chiral
interaction between "chiral centers" on one molecule with a 
center of local anisotropic polarizability on another molecule.
However, this same characterization does {\it not} apply to the
mechanism involving local permanent atomic dipole moments.  The dipolar
mechanism leads to an intrinsically three--point chiral interaction
of a type which, as far as we know, has not yet been proposed.
It would be interesting to observe such an interaction for
helical molecules which have a local radial dipole moment.

\vspace{0.2 in} \noindent
5)  Our calculations can potentially be generalized in several directions.
For instance, there seems to be no reason why our results can not be
taken over immediately to discuss the interaction between
flexible polymers.  There the average over spinning (within
a tube surrounding the convoluted polymer shape) can still be
taken.  Then in Eq. (\ref{K2RESULT}) one would replace ${\bf a}_I$
by its local value at atom $i$.  Our calculations can also be
applied to liquid crystal systems containing a mixture of
chiral and achiral molecules.  There one has two types of interactions
to consider.  The first of these is the interaction between adjacent
chiral and achiral molecules to which the results of this paper apply
directly.  The second is the interaction between more widely separated
pairs of chiral molecules.  For this interaction, our result for
$\kappa_{IJ}$ ought to be multiplied by $\epsilon ^{-2}$, where
$\epsilon$ is the static dielectric constant. 

\vspace{0.2 in} \noindent
{\bf Acknowledgements}  We thank our colleagues R. D. Kamien,
E. Mele, P. Nelson, and D. Weitz for helpful interactions.  This work
was supported in part by the National Science Foundation under
Grant No. DMR 95-20175 (ABH) and DMR 97-30405 (TCL).

%%%%%%%%%%%%%%%%%%%%%%%%%%%%%%%%%%%%%%%%%%%%%%%%%%%%%%%%%%%%%%%%%%%%
%%%%%%%%%%%%%%%%%%%%%%%%%%%%%%%%%%%%%%%%%%%%%%%%%%%%%%%%%%%%%%%%%%%%
\newpage
\appendix
\section{Quantum and Classical Averaging}
\label{BLURB}
The energy of interaction of molecules $I$ and $J$ averaged
over their rotational motion when
expressed in terms of a multipole expansion is of the form
\begin{eqnarray}
\label{TENSOR}
[ U_{IJ} ]_{\rm av} &=& \sum _{M \{\alpha\}, N, \{ \beta \} }
\Lambda_{\alpha_1, \alpha_2 , \dots \alpha_n; \beta_1 , \beta_2 , \dots
\beta_m}  (I,J) [ M_{\alpha_1 , \alpha_2, \dots \alpha_n} (\rho_I) ]_{\rm av}
[ N_{\beta_1 , \beta_2 , \dots \beta_m } (\rho_J) ]_{\rm av}\ ,
\end{eqnarray}
where $[ \ \ ]_{\rm av}$ indicates an average over orientations and
${\bf M}$ and ${\bf N}$ are tensors of arbitrary rank
which are functionals of a density on the molecule in question.
For classical two--body interactions these tensors are
multipole moments of the form
\begin{eqnarray}
M_{\alpha_1 , \alpha_2 , \dots \alpha_n} (\rho_I) = 
\int d {\bf r} \rho_I ({\bf r}) r_{\alpha_1} r_{\alpha_2 } 
\dots r_{\alpha_n} \ .
\end{eqnarray}
For classical two--body interactions these tensors are thus
linear functions of the density, so that the orientational average
of the tensor is the same as the tensor evaluated for the
orientationally averaged density:
\begin{eqnarray}
\label{SPIN}
[M_{\alpha_1 , \alpha_2 , \dots \alpha_n} (\rho) ]_{\rm av} =
M_{\alpha_1 , \alpha_2, \dots \alpha_n} ([\rho]_{\rm av} ) \ .
\end{eqnarray}
This means that classically the interaction averaged over the
orientational motion of molecule $I$, say, is the same as the
interaction would be for a molecule having the average (over
orientations) shape.  Thus, classically, spinning a chiral
molecule leads to two--body interactions characteristic of a
uniaxial, i. e. achiral molecule.  Quantum mechanically, the
situation is different, because in second order perturbation
theory the tensor ${\bf M}$, say, in Eq. (\ref{TENSOR}) is a
bilinear function of the density $\rho(I)$ of the form
\begin{eqnarray}
M_{\alpha_1, \alpha_2, \dots \alpha_n} =
\int \rho({\bf r}) d{\bf r} \int d{\bf r}' \rho({\bf r}')
T({\bf r}, {\bf r}') 
r_{\alpha_1} \dots r_{\alpha_k} r'_{\alpha_{k+1}} \dots
r'_{\alpha_n} \ ,
\end{eqnarray}
where $T({\bf r}, {\bf r}')$ depends on the spatial correlations of
the important excited states, and Eq.  (\ref{SPIN}) is incorrect.
In other words, the nonlinear
fluctuation of the electric field of a molecule due to
quantum fluctuations has a chiral component that survives an
average over rotations and thereby distinguishes
between right--handed and left--handed molecules.

\section{Biaxial Orientational Correlations}
\label{MOON}
\begin{figure}
\psfig{figure=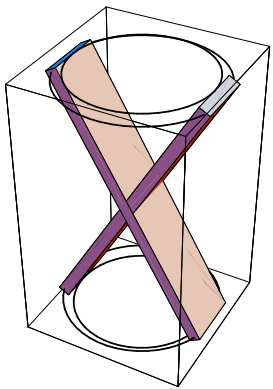}

\vspace{0.2 in} \noindent
\caption{Distribution for which $\gamma-\beta$ is fixed.}
\label{BIAXIAL}
\end{figure}

If the Euler angles are taken to represent the orientation of the
molecule with reference to axes fixed in space such that the $z$--axis
coincides with the axis of nematic order, then the probability
distribution for the orientation of a single molecule must be
independent of $\alpha$.  If the distribution is also independent of
$\gamma$, then it means that for each value of $\beta$, the molecule
spins with equal probability through all angles about its long axis.
However, if we have correlations between $\beta$ and $\gamma$, we can
have a distribution like that describing the orientation of the moon
in which $\gamma-\beta$ assumes a fixed value.  For a molecule, this
distribution is depicted in Fig. \ref{BIAXIAL}.

\section{ORIENTATIONAL AVERAGES}
\label{BCAVE}

In this appendix we evaluate the orientational averages 
(indicated by brackets, $[ \ \ \ ]_{\rm av}$) of the terms
in Eq. (\ref{ETWOEQ}).  In this calculation, we should keep in mind
that we only need keep terms which include one antisymmetric
tensor.  Also only averages of even numbers of powers of components
of $\rhov_i$ are nonzero.  Finally, terms obtained by interchanging
the indices $i$ and $j$ (labeling atoms on different molecules)
can be included implicitly.  With these understandings we use
Eq. (\ref{AVEEQ}) to write
\begin{eqnarray}
T_1 & \equiv & [ (\rhov_{ij} \cdot {\bf D} D^{-3} )_{n0}
( \rhov_{ij} \cdot {\bf D} D^{-3} )_{0n} ]_{\rm av} \nonumber \\
&=&  2 [ (\rho_{i\alpha } D_\alpha D^{-3})_{n0}
( \rho_{i\beta } D_\beta D^{-3} )_{0n} ]_{\rm av} \nonumber \\
&=& \epsilon_{\alpha \beta \gamma} a_{I \gamma} \epsilon_{\mu \nu z}
(r'_{i\mu} D_\alpha D^{-3} )_{n0} (r'_{i\nu} D_\beta D^{-3})_{0n} \nonumber \\ 
&=& \left[ {\bf R} \times {\bf a_I} \cdot {\bf a}_J \right] \Biggl[
(r'_{i\mu } z_j D^{-3} )_{n0} (r'_{i\nu} D^{-3})_{0n}
- (r'_{i \mu } D^{-3} )_{n0} (r'_{i\nu} z_j D^{-3})_{0n} \Biggr]
\epsilon_{\mu \nu z} \nonumber \\
&=& 2 \left[ {\bf R} \times {\bf a_I} \cdot {\bf a}_J \right]
(r'_{i\mu } z'_j D^{-3} )_{n0} (r'_{i\nu} D^{-3})_{0n} \epsilon_{\mu \nu z} \ .
\end{eqnarray}

In terms involving four powers of transverse components, contributions
at the order in $R_{IJ}^{-1}$ which we need require that two
components refer to atom $i$ and two to atom $j$.  Thus
\begin{eqnarray}
T_2 & \equiv & - \case 3/2 \left[ ( [\rhov_{ij}\cdot {\bf D}]^2 D^{-5} )_{n0}
( \rho_{ij}^2 D^{-3} )_{0n} \right]_{\rm av} \nonumber \\
&=& -3 \left[ ( [\rhov_i \cdot {\bf D} ]^2 D^{-5})_{n0}
(\rho_j^2 D^{-3})_{0n} \right]_{\rm av}
- 6 \left[ ( [\rhov_i \cdot {\bf D} ] [\rhov_j \cdot {\bf D}])_{n0}
( [\rhov_i \cdot \rhov_j ] D^{-3})_{0n} \right]_{\rm av} \ .
\end{eqnarray}
The first term gives rise to no antisymmetric terms and can be dropped.
The second term leads to
\begin{eqnarray}
T_2 & \equiv & -3 \left[ (r'_{i\mu} D_\alpha \rho_{j \beta} D_\beta D^{-5}
)_{n0} (r'_{i\mu} \rho_{j\alpha} D^{-3} )_{0n} \right]_{\rm av} \nonumber \\
&& + 3 \left[ (r'_{i\mu} D_\alpha a_{I\alpha} \rho_{j \beta} D_\beta D^{-5}
)_{n0} (r'_{i\mu} a_{I \gamma} \rho_{j\gamma} D^{-3} )_{0n} \right.
\nonumber
\\ && - 3 \left. \epsilon_{\mu \nu z} \epsilon_{\alpha \beta \rho} a_{I\rho}
\langle (r'_{i\mu} D_\alpha \rho_{j\beta} D_\beta D^{-5} )_{n0}
(r'_{i\nu} \rho_{j \gamma } D^{-3} )_{0n} \right]_{\rm av} \ .
\end{eqnarray}
The first term gives zero antisymmetric contribution.  The second and third
terms give identical contributions.  So
\begin{eqnarray}
T_2 &=& 3 \left[ {\bf R} \times {\bf a_I} \cdot {\bf a}_J \right]
(r'_{i\mu} r'_{j\nu} [{\bf D} \cdot {\bf a}_I ] D^{-5})_{n0}
(r'_{i\mu} r'_{j\tau} D^{-3} )_{0n} \epsilon_{\nu \tau z} \ .
\end{eqnarray}

Likewise, keeping only relevant terms, we write
\begin{eqnarray}
T_3 & \equiv & \case 9/4 \left[ ( [ \rhov_{ij} \cdot {\bf D} ]^2 D^{-5} )_{n0}
( [ \rhov_{ij} \cdot {\bf D} ]^2 D^{-5} )_{0n} \right]_{\rm av} \nonumber \\
& = & \left[ \case 9/2 ( [ \rhov_i \cdot {\bf D} ]^2 D^{-5} )_{n0}
( [ \rhov_j \cdot {\bf D} ]^2 D^{-5} )_{0n} \right. \nonumber \\
&& \left. \ +
9 ( [ \rhov_i \cdot {\bf D} ] [\rhov_j \cdot {\bf D} ] D^{-5} )_{n0}
( [ \rhov_i \cdot {\bf D} ] [\rhov_j \cdot {\bf D} ] D^{-5} )_{0n}
\right]_{\rm av} \ .
\end{eqnarray}
The first term leads to zero antisymmetric contribution.  In the second
term there are two equal contributions, one from taking the
antisymmetric term in the average over $\rhov_i$, the other from the
antisymmetric term in the average over $\rhov_j$.  So we write
\begin{eqnarray}
T_3 & = & 9 \epsilon_{\mu \nu z} \left[ ( r'_{i\mu} \rho_{j\beta}
D_\alpha D_\beta D^{-5} )_{n0} (r'_{i\nu} \rho_{j\delta} D_\gamma D_\delta
D^{-5} )_{0n} \epsilon_{\alpha \gamma \rho} a_{I\rho} \nonumber
\right]_{\rm av} \\ &=&
\case 9/2 \epsilon_{\mu \nu z} (r'_{i\mu} r'_{j\tau} D_\alpha D_\beta D^{-5}
)_{n0} (r'_{i\nu} r'_{j\tau} D_\gamma D_\delta D^{-5} )_{0n}
\epsilon_{\alpha \gamma \rho } a_{I\rho} (\delta_{\beta \delta} -
a_{J\beta} a_{J\delta} ) \ .
\end{eqnarray}
Then, using the symmetry between the two matrix elements, we have
\begin{eqnarray}
T_3 &=& 9 \left[ {\bf R} \times {\bf a_I} \cdot {\bf a}_J \right]
(r'_{i\mu } r'_{j \nu} D_\alpha D^{-5})_{n0} (r'_{i\mu} r'_{j\tau}
D_\gamma D_\delta z'_i D^{-5} )_{0n} \epsilon_{\nu \tau z}
(\delta_{\alpha \gamma} - a_{I\alpha} a_{I\gamma} ) \ .
\end{eqnarray}
We set $z'_i = \overline z'_i + \Delta z'_i$.
The term in $\overline z'_i$ vanishes.  Thus
\begin{eqnarray}
T_3 &=& 9 {\bf R} \times {\bf a_I} \cdot {\bf a}_J \overline r'_{i\mu}
(r'_{i\mu } r'_{j \nu} D_\alpha D^{-5})_{n0} (r'_{j\tau}
D_\gamma D_\delta (\Delta z'_i) D^{-5} )_{0n} \epsilon_{\nu \tau z}
(\delta_{\alpha \gamma} - a_{I\alpha} a_{I\gamma} ) \ .
\end{eqnarray}
The matrix elements are symmetric functions of $\mu$ and $\tau$.  So
the antisymmetry of the $\epsilon$ tensor causes this term to vanish.
At higher order in $R_{IJ}^{-1}$ there would be nonzero contributions
from this term. But at the order we consider there are none.

The remaining terms in Eq. (\ref{ETWOEQ}) vanish for reasons
similar to those which made $T_3$ vanish.  So the only contributions
that survive are those written in Eq. (\ref{KAPPAEQ}).

\section{Nonlocal Effects}
\label{LOCAL}
In this appendix we discuss nonlocal corrections contained
in Eq. (\ref{KAPPAEQ}) from terms where $i \not= i'$ and/or
$j \not= j'$.  Rather than give a general argument, we
will illustrate the nature of the argument by considering
specifically the nonlocal corrections to the first term in
Eq. (\ref{KAPPAEQ}).  For this purpose we assume that
the ``unperturbed" energies $E_{n_i n_j}$ can be obtained
from a Hamiltonian of the form
\begin{eqnarray}
\label{PERTURB}
{\cal H} & = & {\cal H}_0 + V_{\rm hop} \nonumber \\
& \equiv & {\cal H}_0 + \sum_{i,j;\alpha ,\beta }
|i \alpha \rangle t_{ij}^{\alpha \beta} \langle j \beta | \ ,
\end{eqnarray}
where ${\cal H}_0$ is completely local:
\begin{eqnarray}
{\cal H}_0 = \sum_{i \alpha} 
|i \alpha \rangle E_i^\alpha \langle i \alpha | \ .
\end{eqnarray}
We assume the states to be strongly localized so that
$|t_{ij}^{\alpha \beta}| \ll E_i^\alpha$ for all indices.

Now we consider the contribution, $T_0$, to $\kappa_{IJ} =
\case 1/2 [\tilde \kappa_{IJ} + \tilde \kappa_{JI} ]$ from the
first line of Eq. (\ref{KAPPAEQ}).  Thus we write
\begin{eqnarray}
T_0 &=&  - e^4 \Biggl[
\langle 0 |  r_{i\alpha} D_{ij}^{-3} {1 \over {\cal E}}
r_{i'\beta} z_{j'} D_{i'j'}^{-3} |0\rangle
- \langle 0 |  r_{i\alpha} z_j D_{ij}^{-3} {1 \over {\cal E}}
r_{i'\beta} D_{i'j'}^{-3} |0\rangle \Biggr] \epsilon_{\alpha \beta z} \ .
\end{eqnarray}
In this appendix all coordinates are taken relative to axes
fixed in the molecule.  Thus, $r_i^\mu$ here denotes what we
called $({\bf r}')_i^\mu$ in the notation of Eq. (\ref{EMBLAZE}).
For simplicity we consider here only the contribution from
virtual states in which both molecules are excited. In that
case, the sums are only over electrons.

Now we expand the matrix elements according to Eq. (\ref{MATELEQ}),
as was done in Eq. (\ref{EXPAND}).  Thereby we get the
corresponding contribution $\delta \kappa_{IJ}$ as
\begin{eqnarray}
\delta \kappa_{IJ}  &=&
\langle 0 | \Delta r_{i \xi} \Delta r_{j \eta} {1 \over {\cal E}}
\Delta r_{i' \sigma} \Delta r_{j \tau} | 0 \rangle \nonumber \\ && \ \
\times \Biggl[ \Biggl( \nabla_{i \xi} \nabla_{j \eta}
\overline r_{i \mu} \overline D_{ij}^{-3} \Biggr)
\Biggl( \nabla_{i' \sigma} \nabla_{j' \tau}
\overline z_{j'} \overline r_{i' \nu} \overline D_{i'j'}^{-3} \Biggr)
\nonumber \\ && \ \ \ \
- \Biggl( \nabla_{i \xi} \nabla_{j \eta}
\overline r_{i \mu} \overline z_j \overline D_{ij}^{-3} \Biggr)
\Biggl( \nabla_{i' \sigma} \nabla_{j' \tau}
\overline r_{i' \nu} \overline D_{i'j'}^{-3} \Biggr) \Biggr] 
\epsilon_{\mu \nu z} \nonumber \\ &=&
\langle 0 | \Delta r_{i \xi} \Delta r_{j \eta} {1 \over {\cal E}}
\Delta r_{i' \sigma} \Delta r_{j \tau} | 0 \rangle \nonumber \\ && \ \
\times \Biggl[ \Biggl( \delta_{\xi \mu} \nabla_{j \eta}
\overline D_{ij}^{-3} + \overline r_{i \mu} \nabla_{i \xi}
\nabla_{j \eta} \overline D_{ij}^{-3} \Biggr)
\Biggl( \delta_{\sigma \nu} \delta_{\tau z} \overline D_{i'j'}^{-3} 
+ \delta_{\sigma \nu} \overline z_j \nabla_{j' \tau}
\overline D_{i'j'}^{-3}
+ \delta_{\tau z} \overline r_{i' \nu} \nabla_{j' \tau} 
\overline D_{i'j'}^{-3} \Biggr) \nonumber \\ && \ \ -
\Biggl( \delta_{\xi \mu} \delta_{\eta z} \overline D_{ij}^{-3}
+ \delta_{\xi \mu} \overline z_j \nabla_{j \eta} \overline D_{ij}^{-3} 
+ \delta_{\eta z} \overline z_j \nabla_{i \xi}
\overline D_{ij}^{-3} \Biggr)
\nonumber \\ && \ \ \ \ \times
\Biggl( \delta_{\sigma \nu} \nabla_{j' \tau} \overline D_{i'j'}^{-3} 
+ \overline r_{i' \nu} \nabla_{i' \sigma} \nabla_{j' \tau}
\overline D_{i'j'}^{-3} \Biggr) \Biggr] \epsilon_{\mu \nu z} \ .
\end{eqnarray}
Here we dropped terms of order $1/R_{IJ}^9$.  In evaluating the
gradients, note that $D_{ij}$ depends on ${\bf r}_i$ (${\bf r}_j$)
only via $z_i$ ($z_j$).  Thus $\nabla_{i\xi}\nabla_{j\eta}
\overline D_{ij}^{-3}$ is only nonzero for $\xi=\eta=z$.

The terms of greatest interest are those of order $1/R_{IJ}^7$, because
such terms are of potentially lower order than the local terms we kept
of order $1/R_{IJ}^8$.  These leading order terms are
\begin{eqnarray}
\delta \kappa_{IJ}  &=&
\langle 0 | \Delta r_{i \mu} \Delta r_{j z} {1 \over {\cal E}}
\Delta r_{i' \nu} \Delta r_{j z} | 0 \rangle 
\left( \overline D_{i'j'}^{-3} \nabla_{jz} \overline D_{ij}^{-3} -
\overline D_{ij}^{-3} \nabla_{j'z} \overline D_{i'j'}^{-3} \right) 
\epsilon_{\mu \nu z} \nonumber \\ & \equiv &
\langle 0 | \Delta r_{i \mu} \Delta r_{j z} {1 \over {\cal E}}
\Delta r_{i' \nu} \Delta r_{j z} | 0 \rangle \Biggl[
f(z_i, z_j, z_{i'}, z_{j'}) -
f(z_{i'}, z_{j'}, z_i, z_j) \Biggr] \ ,
\end{eqnarray}
where $f \sim 1/R_{IJ}^7$.
Note that when the states are localized, i. e. when $i=i'$ and
$j=j'$, the factor in large square brackets vanishes.  Now
consider expanding $\cal E$ as in Eq. (\ref{PERTURB}), so that
\begin{eqnarray} 
{1 \over {\cal E}} = {1 \over E_0 - {\cal H}_0} +
{1 \over E_0 - {\cal H}_0} V_{\rm hop} {1 \over E_0 - {\cal H}_0} +
{1 \over E_0 - {\cal H}_0} V_{\rm hop} {1 \over E_0 - {\cal H}_0}
V_{\rm hop} {1 \over E_0 - {\cal H}_0} + \dots \ ,
\end{eqnarray}
where $E_0 - {\cal H}_0 \sim E$, where $E$, the typical
excitation energy, is much larger than $t$, a typical hopping
matrix element.  This equation implies that when it requires $m$
hops for an electron to move from site $i$ to site $i'$ and  $n$
hops for an electron to move from site $j$ to site $j'$, then the
matrix element will be of relative order $(t/E)^{(m+n)}$.  Thus
\begin{eqnarray}
\Delta f \equiv f(z_i, z_j, z_{i'}, z_{j'}) -
f(z_{i'}, z_{j'}, z_i, z_j) \sim
(t / E ) \nabla f  \sim
(t/E) (z_i - z_{i+1})/R_{IJ}^8 \ .
\end{eqnarray}
We see that the ratio of this nonlocal contribution to the local
contribution of Eq. (\ref{KAPPAEQ}) is of order
$\Delta f /(r_\perp/R_{IJ}^8)$,
where $r_\perp$ is a typical value of $\overline x_i$ or
$\overline y_i$.  This ratio is thus of order
$(t/ E )(z_{i+1}-z_i)/r_\perp$.  Normally
$(z_{i+1}-z_i)/r_\perp$ is of order unity, so indeed the
nonlocal contributions are of relative order $t/E$
and can reasonably be neglected.

\section{Positional Correlations} 
\label{PERP}

In this appendix we consider how energy of interaction for the system
of molecules is effected by the relative distribution of molecules.
A simple way to address this issue is to evaluate the chiral interaction
as a function of $Z_{IJ}\equiv Z$, the $z$--component of ${\bf R}_{IJ}$.
We assume that it suffices to do this for helical molecules, in which case
the calculations can be done explicitly.
Previously we had set $Z=0$ and had considered the contribution
to the torque field from a shell of six neighbors taken to lie in the
equatorial plane.  Here we show numerically that this approximation
is quite reasonable.  We study the dependence of
$S_{IJ}\equiv \sum_{ij} S_{ij}$
on $Z_{IJ}$.  We still assume perfect nematic order, so that
${\bf a}_I = {\bf a}_J = e_z$.  Then the sum in Eq. (\ref{S2EQ}) becomes
\begin{eqnarray}
S_{IJ}(Z) = \sum_{ij} [ R^2 +  (Z +  z_i - z_j)^2]^{-5}
[R^2 - (Z+z_i-z_j)^2] \ ,
\end{eqnarray}
and we see that
\begin{eqnarray}
{\kappa_{IJ}^{(2)}(Z) \over \kappa_{IJ}^{(2)}(0) }
= {S_{IJ}(Z) \over S_{IJ}(0) } \ .
\end{eqnarray}

For the one--molecule terms we similarly note that the
$Z$--dependence in Eq. (\ref{KAPPA1EQ}) is reproduced by writing
\begin{eqnarray}
\tilde \kappa^{(1)}_{IJ}(Z) & \propto & \sum_{i i' j} q_i q_{i'}
(\overline x'_i \overline y'_{i'} - \overline x'_{i'} \overline y'_i)
\frac{ \left( \overline {\bf D}_{i'j} \cdot {\bf a}_J \right) }
{{\overline D}_{ij}^3 {\overline D}_{i'j}^5} \nonumber\\
&=& \sum_{i i' j} 
{ \wp^2_0 (Z+ \overline z'_{i'}- \overline z'_j)
\sin[q (\overline z'_{i'} - \overline z'_i )] \over
[ R^2 + (Z + \overline z'_i - \overline z'_j )^2 ]^{3/2}
[ R^2 + (Z + \overline z'_{i'} - \overline z'_j )^2 ]^{5/2} } \ ,
\end{eqnarray}
in the notation of Eqs. (\ref{KAPPA1EQ}) and (\ref{KAPPA1}).

\begin{figure}[hb]
\psfig{figure=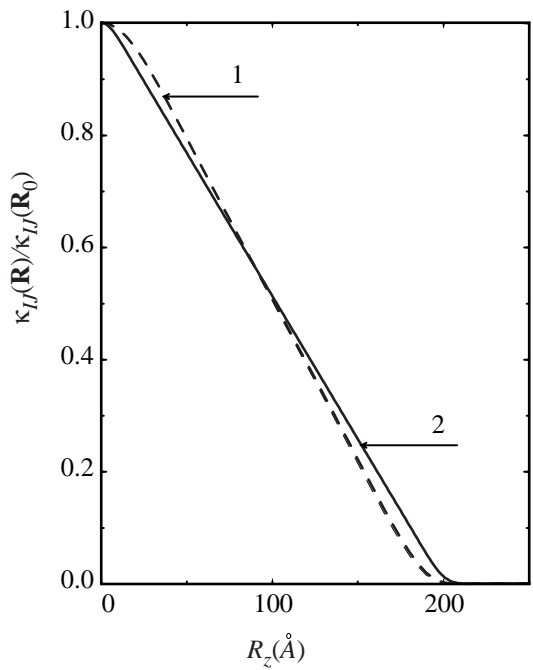}

\vspace{1 in} \noindent
\caption{ The ratio $\kappa_{IJ}^{(n)}(Z)/\kappa_{IJ}^{(n)}(0)$,
where $Z$ is the z-component of ${\bf R}_{IJ}$ for molecules of
length $L= 200 \AA $ and intermolecular separation $R=20 \AA $.
For $n=1$ we show essentially indistinguishable curves for
$q=0.0444 (\AA )^{-1}$ and for $q=0.1333 (\AA )^{-1}$. For
$n=2$ this ratio does not depend on $q$.}
\label{KAPPAFIG}
\end{figure}

These results allow us to compute the ratio
$\kappa^{(n)}_{IJ}(Z) / \kappa^{(n)}_{IJ}(0)$ which is shown
in Fig. \ref{KAPPAFIG} for $n=1$ and $n=2$.
This result is representative of the situation for a wide range
of parameters.  As one might expect, the contribution to the torque
field decreases strongly as $|Z|/L$ increases towards unity.
Accordingly, the approximation of including
only the effect of equatorial neighbors is a good one.

\section{Contributions to {\protect$h^{(1)}$}}
\label{h1app}

In this appendix we discuss the evaluation of the one--molecule
contributions to $\tilde \kappa_{IJ}$.  We consider the terms in
the last two lines of Eq. (\ref{KAPPAEQ}).  We will analyze the
one--molecule contributions which arise when $i=i'$ but $j$ and $j'$ 
are in general different.  In the intermediate excited state
only atom $i$ is in an excited state.  Atoms $j$ and $j'$ remain
in their ground states.  Calling this term $T$ we write
\begin{eqnarray}
T &=& - 3 \sum_{i,j,j'} e^2 q_j q_{j'} \langle 0 | x'_i y'_{j'} D_{ij'}^{-3}
{1 \over {\cal E}} x'_i x'_j {\bf D}_{ij} \cdot {\bf a}_I D_{ij}^{-5} |0
\rangle + \dots \ ,
\end{eqnarray}
where the dots denote the three additional terms required to make the
expression be rotationally invariant.  (These can be reconstructed at the
end of the calculation.)  Using the expansion of Eq. (\ref{MATELEQ}) we have
\begin{eqnarray}
T&=& - 3 e^2 \sum_{i j j'} q_j q_{j'} \langle 0 |
[ \Delta x'_i \overline D_{ij'}^{-3} + \overline x'_i \Delta z'_i
(\nabla_{iz} D_{ij'}^{-3} ) ] \overline y'_{j'} {1 \over {\cal E}} \nonumber \\
&& \times \left[ \Delta x'_i {\bf D}_{ij} \cdot {\bf a}_I D_{ij}^{-5}
+ \overline x'_i \Delta z'_i \nabla_{iz} [ ({\bf D}_{ij}\cdot {\bf a}_I)
D_{ij}^{-5} ] \right] \overline x'_j | 0 \rangle + \dots \nonumber \\
&\approx& - 3 e^2 \sum_{ijj'} q_j q_{j'} \langle 0| \Delta x'_i
{1 \over {\cal E}} \Delta x'_i | 0 \rangle \overline x'_j \overline y'_{j'}
\overline D_{ij'}^{-3} (\overline {\bf D}_{ij} \cdot {\bf a}_I)
\overline D_{ij}^{-5} + \dots  \ .
\end{eqnarray}
Now we carry the sum over $j$ ($j'$) over the charges that comprise the
dipole moment ${\bf p}_j$ (${\bf p}_{j'}$) on atom $j$ ($j'$) to get
\begin{eqnarray}
T &=& - 3 e^2 \sum_{ijj'} \langle 0| \Delta x'_i
{1 \over {\cal E}} \Delta x'_i | 0 \rangle 
[ p'_{yj'} \overline D_{ij'}^{-3} + p'_{zj'} \overline y'_{j'} ( \nabla_{j'z}
D_{ij'}^{-3} ) ] \nonumber \\ && \ \ \times
\left[ p'_{xj} (\overline {\bf D}_{ij} \cdot {\bf a}_I) \overline D_{ij}^{-5}
+ p'_{zj} \overline x'_{j'} \nabla_{jz}
[( \overline {\bf D}_{ij} \cdot {\bf a}_I ) D_{ij}^{-5}] \right]
+ \dots \nonumber \\
&\approx& - 3 e^2 \sum_{ijj'} \langle 0| \Delta x'_i
{1 \over {\cal E}} \Delta x'_i | 0 \rangle p'_{yj'} p'_{xj}
\overline D_{ij'}^{-3}
(\overline {\bf D}_{ij} \cdot {\bf a}_I)  D_{ij}^{-5} + \dots \ ,
\end{eqnarray}
where now $j$ and $j'$ refer to atoms.
Restoring the additional terms to preserve rotational invariance we obtain
\begin{eqnarray}
T&=& - 3 e^2 \sum_{ijj'} [ 
\langle 0| \Delta x'_i {1 \over {\cal E}} \Delta x'_i | 0 \rangle 
+ \langle 0| \Delta y'_i {1 \over {\cal E}} \Delta y'_i | 0 \rangle ]
[ p'_{yj'} p'_{xj} - p'_{xj'} p'_{yj}] \overline D_{ij'}^{-3}
(\overline {\bf D}_{ij} \cdot {\bf a}_I )  D_{ij}^{-5} \ .
\end{eqnarray}
When the indices are relabeled, this result reproduces part of Eq.
(\ref{KAPPA1EQ}).

\section{Evaluation of Integrals in Sec. V}
\label{X}
In this appendix we evaluate the integral $X_1$ in Eq. (\ref{X1EQ})
and an integral needed to evaluate the dipolar interaction energy
between two long helices.

Consider the asymptotic evaluation of Eq. (\ref{X1EQ}),
firstly, in the limit $L \gg R$.  End effects can be shown to be negligible,
in which case the final summation (over $\tilde s_j$) introduces a
factor of $N$ and one sets $s_j=0$.  Also we consider only the
continuum limit in which the sums are replaced by integrals.  One
can show that correct to leading order in $\tilde L^{-1}$, the limits
on the integrals can be extended to $\pm \infty$.  Thus we have the
asymptotic result
\begin{eqnarray}
X_1( \tilde q, \tilde L ) \sim 3N^3 \int_{-\infty}^\infty ds
\int_{-\infty}^\infty ds'  
\Biggl[ {1 - \tilde L^2 ss' \over (1+ \tilde L^2 s^2)^{5/2}
(1+ \tilde L^2 {s'}^2)^{5/2} } \Biggr] (s'-s) \sin [qL (s'-s)] \ .
\end{eqnarray}
For each of the two factors in the denominator we introduce the representation
\begin{eqnarray}
\label{GAUSS}
p^{-5/2} = {1 \over (3\sqrt {2\pi})}
\int_{-\infty}^\infty x^4 e^{- \case 1/2 px^2} dx \ .
\end{eqnarray}
Then the integrations over $s$ and $s'$ can be done analytically
and eventually one finds that
\begin{eqnarray}
X_1(\tilde q , \tilde L \rightarrow \infty ) = 8 (\rho R)^3
(qR)  I_1^2(qR) \ , \ \
\end{eqnarray}
where $I_1$ is defined in Eq. (\ref{INTEQ}) of the text.

The limit $\tilde L \rightarrow 0$ is trivial.  We find that
\begin{eqnarray}
X_1(\tilde q , \tilde L \rightarrow 0 ) = N^3 \phi(\case 1/2 qL) \ ,
\end{eqnarray}
where $\phi(x) = - (3/2) (d/dx)[(\sin x)/x]^2$.

Finally we evaluate the dipolar interaction energy $E_{dd}$ between
two long helical molecules, $a$ and $b$, separated by a distance $R$
along the $x$--axis.  We assume that the radius of the helix is much
less than $R$.  In this limit, in terms of the atomic dipole moments
we write
\begin{eqnarray}
E_{dd} &=& \rho^2 \int_{-L/2}^{L/2} dz_a \int_{-L/2}^{L/2} dz_b
\Biggl[ R^2 + z_{ab}^2 \Biggr]^{-3/2} \Biggl\{
\Biggl[  p_x'' \cos (qz_a + \phi_a)
- c\tilde p_y'' \sin (qz_a + \phi_a) \Biggr] \nonumber \\ & & \ \ \times
\Biggl[  p_x'' \cos (qz_b + \phi_b) - c \tilde p_y'' \sin (qz_b + \phi_b)
\Biggr] \Biggl[ 1 - {3 R^2 \over R^2 + z_{ab}^2 } \Biggr] \nonumber \\ && \ \ + 
\Biggl[ p_x'' \sin (qz_a + \phi_a) + c \tilde p_y'' \cos (qz_a + \phi_a)
\Biggr] \nonumber \\ && \ \ \times
\Biggl[ p_x'' \sin (qz_b + \phi_b) + c \tilde p_y'' \cos (qz_b + \phi_b)
\Biggr] \Biggr\} \ ,
\end{eqnarray}
where $\tilde p_y''= p_y''+ aqp_z''$, $\phi_a$ ($\phi_b$) is the angle of
rotation of molecule $a$ ($b$) about its long axis, and $z_{ab}= z_a - z_b$.
Here we did not include terms involving $p'_z$ which either are
independent of both angles $\phi_a$ and $\phi_b$ or vanish
in the limit $L \rightarrow \infty$.  In that limit we only need to
keep terms which depend on $z_{ab}$, in which case we have
\begin{eqnarray}
E_{dd} & = & \case 1/2 \rho^2 \wp_0^2 L \int_{-\infty}^\infty
\cos(qz_{ab} + \phi_{ab} ) \Biggl[
{ 2z_{ab}^2 - R^2 \over (R^2 + z_{ab}^2 )^{5/2} } \Biggr] d z_{ab} \ ,
\end{eqnarray}
where $\phi_{ab}=\phi_a - \phi_b$.  Using Eq. (\ref{GAUSS}) we obtain
the final result
\begin{eqnarray}
E_{dd} & = & - L \rho^2 \wp_0^2 q^2 I_{-1} \cos \phi_{ab}
\equiv - \case 1/2 V_{dd} \cos \phi_{ab}  \ .
\end{eqnarray}

\newpage 

\newpage

\begin{thebibliography}{99}
\bibitem{SIGN}
We introduce the chiral term with opposite sign to the one
taken in Ref. \protect\onlinecite{HKL}, so that positive $h$
corresponds to right--handed chirality.
\bibitem{LOCALIZED}
Long polymeric molecules inevitably are nonuniform.  For such
a one--dimensional system, we expect the states to be strongly
localized by disorder, in which case the band width is no
longer relevant.
\bibitem{MIRROR}
It is clear from Eq. (\protect\ref{SIGMAEQ}) that
$\sigma'$ is zero if there exists a mirror plane
perpendicular to any of the molecule fixed axes.
The expression given in this equation does not vanish
if there is an oblique mirror plane, but such a
situation is incompatible with our assumption that
we have identified $e'_z$ as the axis about which
the molecule spins.
\bibitem{MOON}
This average, reminiscent of the motion of
the moon about the earth, is not properly incorporated by simply
averaging independently over $\beta_I$ and $\gamma_I$.
\bibitem{HANDED}
In general, the handedness of a molecule consisting of discrete atoms
is actually not a precise concept, although it may seem intuitively
obvious for a helix.  If the helix is continuous, then it is
possible to unambiguously distinguish the handedness of the structure.
This question is discussed in more detail by HKL.
\end{thebibliography}
\end{document}